\documentclass[11pt]{article}

\usepackage{fancyhdr}
\usepackage{url}

\usepackage{libertine}
\usepackage{libertinust1math}
\usepackage[T1]{fontenc}

\usepackage{cite}
\usepackage{amsmath,amssymb,amsfonts}

\usepackage{graphicx}
\usepackage{textcomp}
\usepackage{xcolor}

\usepackage{amsmath}
\usepackage{amsthm}
\usepackage{comment}
\usepackage{multicol}
\usepackage{bbm}
\usepackage{bm}
\usepackage{graphicx}
\usepackage{enumerate}

\usepackage{resizegather}

\usepackage{subcaption}
\usepackage{color}

\usepackage{tikz}
\usetikzlibrary{plotmarks}
\usepackage{pgfplots}
\usetikzlibrary{patterns}
\pgfplotsset{compat=1.3}
\usetikzlibrary{backgrounds}
\usepackage{multirow}
\usepackage{tabularx}
\usepackage[breaklinks,colorlinks]{hyperref}
\hypersetup{breaklinks,colorlinks,citecolor=blue,linkcolor=blue}
\usepackage{threeparttable}

\usepackage[noend]{algpseudocode}
\usepackage{algorithm}

\makeatletter
\def\thmheadbrackets#1#2#3{%
	\thmname{#1}\thmnumber{\@ifnotempty{#1}{ }\@upn{#2}}%
	\thmnote{ {\the\thm@notefont[#3]}}}
\makeatother

\newtheoremstyle{myDefinitionStyle}
{}
{}
{\itshape}
{}
{%
	\bfseries
}
{.}
{ }
{\thmname{#1}\thmnumber{ #2}\thmnote{ (#3)}}%

\newtheoremstyle{myAlgorithmStyle}
{}
{}
{}
{}
{%
	\bfseries
}
{.\\}
{ }
{\thmname{#1}\thmnumber{ #2}\thmnote{ (#3)}}%

\newcommand{\PP}[1]{
	\vspace{2px}
	\noindent{\bf {#1}{.}}
}
\newcommand{\PC}[1]{
	\vspace{2px}
	\noindent{\bf \IfEndWith{#1}{:}{#1}{#1:}}
}

\newtheoremstyle{myTheoremStyle}
{}
{}
{\itshape}
{}
{%
	\bfseries
}
{.}
{ }
{\thmname{#1}\thmnumber{ #2}\thmnote{ (#3)}}%

\theoremstyle{myDefinitionStyle}
\newtheorem{Definition}{Definition}

\theoremstyle{myTheoremStyle}
\newtheorem{Theorem}{Theorem}

\newtheorem{Protocol}{Protocol}
\newtheorem{Simulator}{Simulator}

\theoremstyle{myAlgorithmStyle}
\newtheorem{Alg}{Algorithm} 
\usepackage{amsthm}

\usepackage{enumitem}
\setlist[itemize]{leftmargin=*,label=\textbullet}
\setlist[enumerate]{leftmargin=*}

\usepackage{mdframed}

\def\Snospace~{\S{}}

\def\equationautorefname~#1\null{(#1)\null}
\def\Equationautorefname~#1\null{(#1)\null}

\pgfplotsset{
	tick label style = {font=\scriptsize},
	every axis label = {font=\scriptsize},
	legend style = {font=\scriptsize},
	label style = {font=\scriptsize}
}

\newcommand{\revnew}[1]{{{#1}}}

\newcommand{\Setup}{\ensuremath{\mathsf{Setup}}}
\newcommand{\pp}{\ensuremath{\mathsf{pp}}}

\newcommand{\Prover}{\ensuremath{\mathcal{P}}}
\newcommand{\Verifier}{\ensuremath{\mathcal{V}}}
\newcommand{\Extor}{\ensuremath{\mathcal{E}}}
\newcommand{\Adv}{\ensuremath{\mathcal{A}}}
\newcommand{\Sim}{\ensuremath{\mathcal{S}}}

\newcommand{\secparam}{\ensuremath{\lambda}}

\newcommand{\cm}{\ensuremath{\mathsf{cm}}}

\newcommand{\Infer}{\ensuremath{\Prover}}

\newcommand{\Verify}{\ensuremath{\Verifier}}

\newcommand{\Commit}{\ensuremath{\mathsf{Com}}}

\newcommand{\Eval}{\ensuremath{\mathsf{Eval}}}
\newcommand{\VerifyEval}{\ensuremath{\mathsf{Verify}}}

\renewcommand{\PC}{\ensuremath{\mathsf{zkPC}}}

\newcommand{\mlparam}{\ensuremath{\mathbf{w}}}

\newcommand{\F}{\ensuremath{\mathbb{F}}}
\newcommand{\rel}{\ensuremath{\mathcal{R}}}

\newcommand{\Gen}{\ensuremath{\mathcal{G}}}

\newcommand{\mat}[1]{\ensuremath{\mathbf{#1}}}

\newcommand{\zkp}{\ensuremath{\mathsf{zkp}}}

\newcommand{\zkmlip}{\ensuremath{\mathsf{zkMLIP}}}

\newcommand{\funcml}{\ensuremath{\mathcal{F}_{\mathsf{mlip}}}}

\newcommand{\sys}{\ensuremath{\mathsf{ezDPS}}}

\newcommand{\vect}[1]{\ensuremath{\mathbf{#1}}}

\newcommand{\set}[1]{\ensuremath{\mathcal{#1}}}

\DeclareMathOperator*{\argmax}{argmax} 

\usepackage{etoolbox}
\newcommand{\circled}[2][]{%
	\tikz[baseline=(char.base)]{%
		\node[shape = circle, draw, fill=red, color=red, inner sep = .4pt]
		(char) {\phantom{\ifblank{#1}{#2}{#1}}};%
		\node at (char.center) {\makebox[0pt][c]{\color{white}{#2}}};}}
\robustify{\circled}

\makeatletter
\def\blfootnote{\gdef\@thefnmark{}\@footnotetext}
\makeatother
\date{}

\title{{\textsf{ezDPS}: An Efficient and Zero-Knowledge Machine Learning Inference Pipeline}\thanks{This paper is to appear in Privacy-Enhancing Technologies Symposium (PETS) 2023.}}

\author{
	Haodi Wang\thanks{{Beijing Normal University / Virginia Tech, \href{mailto:whd@mail.bnu.edu.cn}{whd@mail.bnu.edu.cn}.}}
	\and 
	Thang Hoang\thanks{Virginia Tech, \href{mailto:thanghoang@vt.edu}{thanghoang@vt.edu}.} 
}

\begin{document}
  \maketitle

\begin{abstract}
	Machine Learning as a service (MLaaS) permits 
resource-limited clients to access powerful data analytics services ubiquitously. 
Despite its merits, 
MLaaS poses significant concerns 
regarding the integrity of delegated computation and the privacy of the server's model parameters.
To address this issue, Zhang et al. (CCS'20) initiated the study of zero-knowledge Machine Learning (zkML). 
Few zkML schemes have been proposed afterward; 
however, they focus on sole ML classification algorithms that may not offer satisfactory accuracy or require large-scale training data and model parameters,
which may not be desirable for some applications.

We propose $\sys$, a new efficient and zero-knowledge ML inference scheme. 
Unlike prior works, \sys~is a zkML pipeline in which the data is processed in multiple stages for high accuracy. 
Each stage of \sys~is harnessed with an established ML algorithm that is shown to be effective 
in various applications, including Discrete Wavelet Transformation, 
Principal Components Analysis, 
and Support Vector Machine.
We design new gadgets to prove ML operations effectively. 
We fully implemented \sys~and assessed its performance on real datasets. 
Experimental results showed that 
\sys~achieves one-to-three orders of magnitude more efficient than 
the generic circuit-based approach in all metrics 
while maintaining more desirable accuracy than single ML classification approaches.

\end{abstract}

\section{Introduction}

Machine learning (ML) has grown to become a game-changer for the humane society.
A well-trained ML model can effectively aid in performing highly complicated tasks 
such as medical diagnosis, natural language processing, intrusion detection, or financial forecasting.
However, since a powerful ML model requires a large amount of data and computational resources for training, it may not be widely accessible to individuals or small organizations.
To address this issue, 
Machine Learning as a Service (MLaaS) has been proposed, which permits resource-limited clients to access useful ML services (e.g., visualization, training, classification) offered by cloud providers.

Despite its usefulness, MLaaS has posed new integrity and privacy concerns.
When the client delegates the ML computation to the MLaaS server, 
it is not clear if she will receive a reliable response.
A corrupted server may process the client data arbitrarily or even substitute it with malicious data, 
making the outcome untrustworthy.
This is especially critical for sensitive applications 
such as medical diagnosis, intrusion detection, or fraud detection.
Computation integrity can be addressed with Verifiable Computation (VC), 
in which the MLaaS server attaches a \emph{proof} 
to show that the computation is carried out correctly \cite{ghodsi2017safetynets}.
However, VC itself may not be sufficient for MLaaS because 
it only enables computation integrity but not the privacy of the parameters used in the computation.
In MLaaS, the server uses its private ML model to process the client data.
This sophisticated model may cost significant resources to obtain and, therefore, 
it is considered the intellectual property of the server.
Moreover, such models may also be trained from sensitive training data (e.g., medical).
As a result, it is undesirable that the MLaaS server
 leak any information about its private ML models when processing the client query.

The above privacy concern in MLaaS can be addressed by adding the \emph{zero-knowledge} property to the VC proof, 
which permits verifiable computation without leaking any information other than the computation result \cite{goldwasser1989knowledge}.
Preliminary zero-knowledge VC (zkVC) protocols are computation and communication expensive with strong assumptions.
Thanks to the recent advancements in cryptography, recent zkVC protocols have become more practical. 
Recently, Zhang et al. \cite{zhang2020zero} have initiated zero-knowledge ML (zkML) research.
In zkML inference, the server first commits to its ML model parameters 
and then provides an interface for the client to process her data sample.
Given a client data sample, 
the server returns the ML computation result along with a zero-knowledge proof, 
which permits the client to verify the ML computation regarding the committed model without learning the model parameters in the proof.
%

Few zkML schemes have been proposed such as zero-knowledge Decision Tree (zkDT) \cite{zhang2020zero}, and zero-knowledge deep learning \cite{lee2020vcnn,liu2021zkcnn}.
%
%
Although the decision tree (DT) is simple with the lightweight model parameters, 
it offers limited accuracy for predicted outcomes.
Deep neural networks (DNNs) permit a high accuracy rate, 
however, it may require a large amount of training data and heavy model parameters and, therefore, may not be ideal for some applications.
Most zkML schemes (e.g., \cite{lee2020vcnn,liu2021zkcnn, zhang2020zero}) also focus solely on the final ML inference phase, while the data is generally processed via a so-called ML pipeline
with multiple processing phases 
(e.g., (pre)processing, feature extraction, and classification) to achieve a desirable performance.
%
%
%
%
Thus, there is a need to develop a zero-knowledge ML pipeline to achieve balanced performance and model complexity for some applications.

\PP{Research Objective} {
	The objective of this paper is to design an efficient and zero-knowledge ML pipeline, 
	which permits the data to be processed in multiple phases for accuracy 
	while at the same time, 
	permitting the verifiability without leaking private model parameters 
	at every processing phase.}

\PP{Our Contributions}
In this paper, 
we propose \sys, an efficient and zero-knowledge ML inference pipeline, 
which offers desirable security properties (e.g., zero-knowledge, verifiability) 
along with high accuracy for MLaaS.
$ \sys $ comprises typical phases of an ML pipeline, 
including data (pre)processing, feature extraction, and classification.
In $ \sys $, we instantiate with established classical ML algorithms including, 
Discrete Wavelet Transformation (DWT) \cite{vonesch2007generalized} for preprocessing,
Principal Components Analysis (PCA) \cite{wold1987principal} for feature extraction,
and  Support Vector Machines (SVM) \cite{2003Radius} for classification due to their popularity and wide adoption in many applications \cite{SVM:PCA:DWT:martis2013ecg,SVM:PCA:DWT:luz2016ecg}.
%
%
To our knowledge,
we are the first to propose a zero-knowledge ML inference pipeline.
Our concrete contributions are as follows.

\begin{itemize}

	\item \PP{New gadgets for critical ML operations} 
	We create new gadgets for proving essential ML operations in arithmetic circuits such as exponentiation, absolute value, 
	and max/min in an array  (\autoref{sec:function}).
	These gadgets are necessary for proving concrete ML algorithms in our proposed scheme but also for other ML operations such as deep learning.
	%
	\item \PP{New zero-knowledge ML inference pipeline scheme with high accuracy}
	Built on top of our proposed gadgets, 
	we design \sys, an efficient and zero-knowledge ML inference pipeline, permits the data to be processed with effective ML algorithms for high accuracy (\autoref{sec:ml:ml ins}).
	We design new methods to prove DWT, PCA, and multi-class SVM with different kernel functions via an optimal set of arithmetic constraints. 
	\sys~significantly outperforms the generic approaches both in asymptotic and concrete performance metrics.
	\sys~is designed to be compatible with any zkVC backend (similar to \cite{zhang2020zero}), thus, its concrete efficiency can be further improved when adopted with a more efficient zkVC.
	We also propose a zero-knowledge proof-of-accuracy scheme to enable public validation of the effectiveness of the committed ML model on public datasets (\autoref{sec:zkPoA}). 
	
	\item \PP{Formal security analysis} 
	We present a formal security model for zero-knowledge ML inference pipeline (\autoref{sec:models}) and rigorously analyze the security of our scheme.
	We prove that $ \sys $ satisfies the security of a zero-knowledge ML inference pipeline (\autoref{sec:analysis}).

	\item \PP{Full-fledged implementation, evaluation, and comparison} 
	We fully implemented our proposed techniques (\autoref{sec:impl}) and conducted a comprehensive experiment to evaluate their performance in real-world environments. (\autoref{sec:exp}).
	Experiments on real datasets showed that \sys~achieves {one-to-three} orders of magnitude more efficient than the generic circuit approaches in {all} performance metrics (i.e., proving time, verification time, proof size).
	%
	%
	Our implementation is available  at
	
	\begin{center}
		\href{https://github.com/vt-asaplab/ezDPS}{https://github.com/vt-asaplab/ezDPS}
	\end{center}
\end{itemize}

\PP{Remark}
In this paper, we focus on the verifiability of the ML inference task
and the privacy of the server model in the integrity proof.
Our technique does not permit client data privacy, 
in which the client sends plaintext data to the server for computation.
This model is different from 
the standard privacy-preserving ML inference (PPMLI) 
(e.g., \cite{juvekar2018gazelle,2017Oblivious,2019XONN,dathathri2019chet,gilad2016cryptonets}), 
which preserves the privacy of the client and server against each other but not computation integrity (see \autoref{sec:relatedwork} for more details). 
To our knowledge, it is not clear how to combine zero-knowledge with PPMLI efficiently
to enable both client and server privacy plus computation integrity. 
We leave such an investigation as our future work.

\PP{Application use-cases}
Our zkML inference scheme can be found useful in various applications. 
First, 
it can be used to enable \emph{proof-of-genuine} ML services, 
in which the service provider can prove that its ML model is of high quality, 
and the inference result is computed from the same model.
Another application is
a fair ML model trading platform with {try-before-buy}, 
in which the buyer can attest to the ML model quality before purchase, while the sellers do not want to reveal their model first.
Finally, our technique can partially address the \emph{reproducibility} problem in ML \cite{ML:reproducibility:heil2021reproducibility}, 
where some ML models are claimed to achieve high accuracy without having a proper way to validate them.
Our technique can offer a solution to this issue, in which the model owner can prove that there exists an ML model that can achieve such accuracy (see \autoref{sec:zkPoA}), 
and the verifier can verify that statement efficiently in zero knowledge.

\section{Preliminaries}
\PP{Notations}For $ n \in \mathbb{N}$, we denote $[1,n] = \{1,\dots,n\}$. Let $ \secparam $ be the security parameter and $ \mathsf{negl}(\cdot) $ be the negligible function.
We denote a finite field as  $\mathbb{F}$. 
PPT stands for Probabilistic Polynomial Time.
We use bold letters, e.g., $\vect{a}$ and $ \mat{A} $,  to denote  vector and matrix, respectively.
$ \mat{A}^\top $ means the transpose of $ \mat{A} $.
We write $ \vect{a} \vect{b}  $ (or $ \vect{a} \cdot \vect{b} $) to denote dot product and $ \mat{A} \circ \mat{B} $ to denote Hadamard (entry-wise) product. 
We use $\stackrel{c}{\approx}$ to denote that two quantities are computationally indistinguishable.

\subsection{Commit-and-Prove Argument Systems}
\label{sec:pre:zk}
\PP{Argument of knowledge} 
An argument of knowledge for an NP relation $ \rel $ is a protocol between a prover $ \Prover $ and a verifier $ \Verifier $, in which $ \Prover $ convinces  $ \Verifier  $ that it \emph{knows} a witness $ w $ for some input in an NP language $ x \in \mathcal{L} $ such that $(x, w) \in \rel$. Let $\langle\Prover, \Verifier\rangle$ denote a pair of PPT interactive algorithms.
A zero-knowledge argument of knowledge is a tuple of PPT algorithms $\zkp = ( \Gen, \Prover,\Verifier )$ that satisfies the following properties.

\begin{itemize}
	\item \emph{Completeness. } For any $ (x, w) \in \rel $ and $\pp \gets \Gen(1^\secparam)$, it holds that 
$$	\langle \Prover(w,\pp),\Verifier(\pp)\rangle(x) = 1 $$

	\item \emph{Knowledge soundness. } For any PPT prover $ \Prover^* $, there exists a PPT extractor $ \Extor $ such that given the access to the entire execution process and the randomness of $ \Prover^* $, $ \Extor $ can extract a witness $ w $ such that $\pp \gets \Gen(1^\secparam), \pi^*  \gets \Prover^*(x,\pp), w \gets \Extor^{\Prover^*}(x, \pi^*,\pp)$ and 
	\begin{gather*}
		\Pr \left[
			 (x, w) \notin \rel  \land \mathcal{V}(x, \pi^*,\pp) = 1
		\right] \leq \mathsf{negl}(\secparam)
	\end{gather*}
	\item  \emph{Zero-knowledge. } 
	There exists a PPT simulator $ \Sim $ such that for any PPT algorithm $\Verifier^*$, auxiliary input $ z \in \{0,1\}^*$, $(x,w) \in \rel $, $\pp \gets \Gen(1^\secparam)$:	
	\[
	\mathsf{view}(\langle \Prover(w,\pp), \Verifier^*(z,\pp) \rangle(x)) \stackrel{c}{\approx} \Sim^{\Verifier^*}(x,z)
	\]
	where $	\mathsf{view}(\langle \cdot, \cdot \rangle(x))$ denotes the distribution of the transcript of interaction.
	
\end{itemize}

\PP{Commit-and-Prove zero-knowledge proof}
Commit-and-Prove (CP) Zero-Knowledge Proof (ZKP) permits the prover to prove the NP-statements on the committed witness. 
Most generic ZKP protocols support CP paradigm and the most efficient CP-ZKP protocols 
harness the succinct polynomial commitment scheme (e.g., \cite{kate2010constant}) to achieve succinctness properties.
The prover first commits to the witness $ w $ using a 
zero-knowledge polynomial commitment scheme before proving an NP statement, and the verifier takes the committed value as an additional input for verification.
We denote the commitment algorithm for CP-ZKP as $ \cm_w \gets \zkp.\Commit(w,r, \pp) $, where $ r $ is the randomness chosen by the prover.

In our framework, we use Spartan \cite{setty2020spartan} (with Hyrax \cite{wahby2018doubly} as the underlying polynomial commitment scheme) as the backend \PC-based CP-ZKP protocol due to its succinctness properties 
(e.g., linear proving time,  sublinear verification time, and proof size), transparent setup, and support generic Rank-1 Constraint System (R1CS).
Generally speaking, Spartan supports NP statements expressed as R1CS, which shows that  there exists a vector $z=(x, 1, w)$ such that $\mathbf{A}z \circ \mathbf{B}z = \mathbf{C}z$, where $\mathbf{A}, \mathbf{B},\mathbf{C}$ are matrices for the arithmetic circuits, $x$ is the public input (statement), $ w $ is the witness of the prover. {All the witnesses are encoded as a polynomial on the Lagrange basis.}
Since it is easy to convert arithmetic statements into R1CS,
our main focus is to create arithmetic constraints for proving algorithms in the ML pipeline efficiently
that can be realized with Spartan or any CP-ZKP backend.

\begin{Theorem}[Spartan ZKP \cite{setty2020spartan}]\label{theorem:spartan}
	Let $ \F $ be a finite field and $ \mathcal{C}_\F $ be a family of the arithmetic circuit over $\F$ of size $ n $.
	Under standard cryptographic hardness assumptions,
	there exists a family of succinct argument of knowledge for the relation $$ \rel = \{ (C,x; w) : C \in C_\F \land C(x;w) = 1\}$$
	where $ x $ and $ w $ are the public input and the auxiliary input to the circuit $ C $, respectively,
	and the prover incurs $ O(n) $	to $O(n \log n)$ overhead, the verifier’s time and communication costs range from $O(\log^2 n)$ to $O(\sqrt{n})$ depending on the underlying polynomial commitment schemes being used for multilinear polynomials.
\end{Theorem}
{Note that since Spartan is established on the polynomial commitment schemes, it can support CP-ZKP paradigm.}

\vspace{-.9em}
\subsection{Machine Learning Pipeline}
\begin{figure}[t]
	\centering
	\includegraphics[width=0.6\columnwidth]{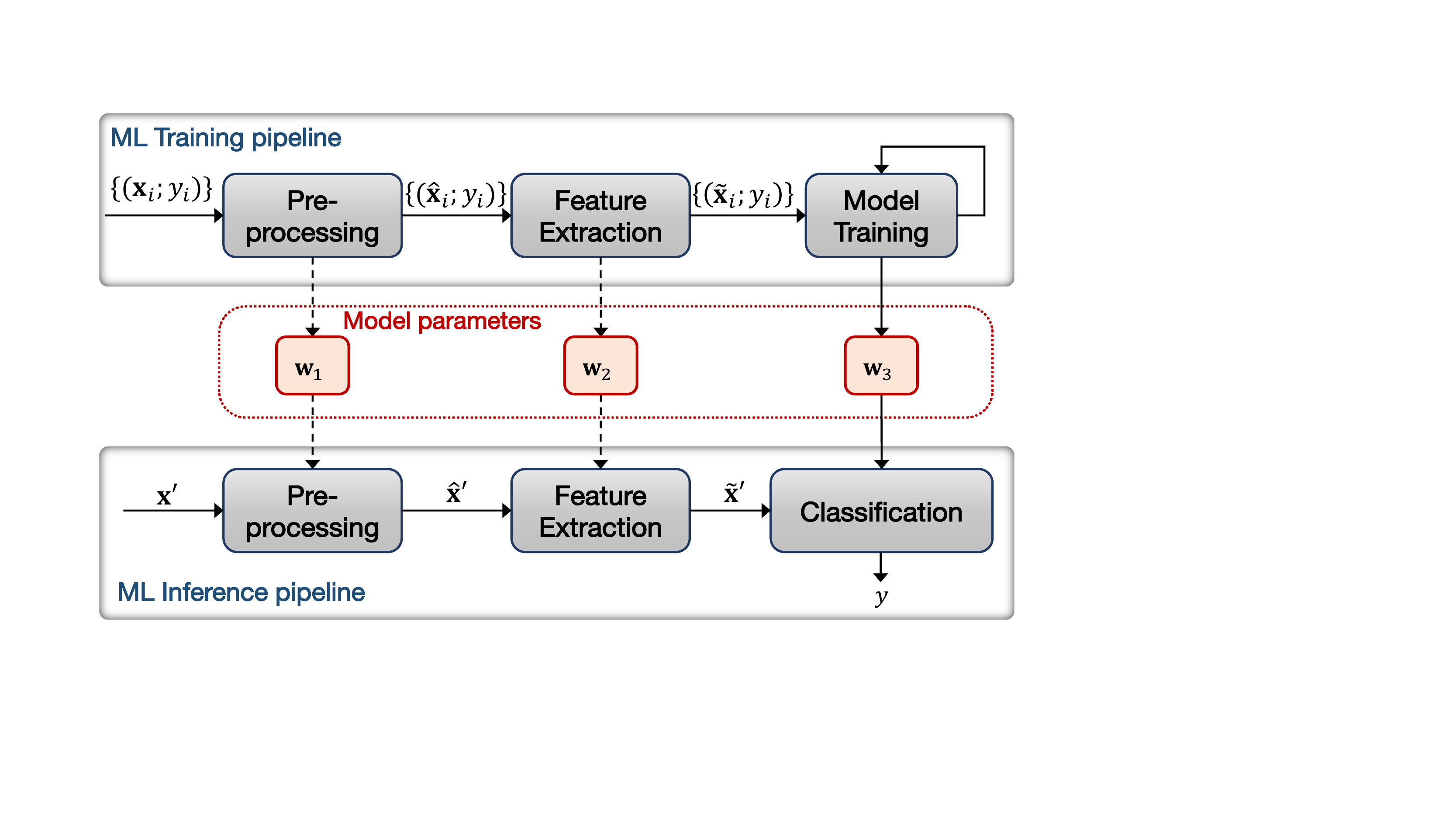}\vspace{-.7em}
	\caption{A general ML pipeline.}
	\label{fig:generalML}\vspace{-.8em}
\end{figure}

ML pipeline is an end-to-end process that consists of multiple data processing phases to train an ML model from a large-scale dataset effectively and to predict an inference result for a new observation accurately \cite{jiang2017learning}.
An effective ML pipeline contains three main phases, including data preprocessing, feature extraction, and ML training/inference as illustrated in  \autoref{fig:generalML}.
In data preprocessing, raw samples $ \vect{x} \in \F^m $ are collected, and then some preprocessing technique is used to reduce the impact of noise in the collection environment.
Feature extraction \revnew{extracts the most prominent dimension of the preprocessed data} so that only a small set of features $ \vect{x}' \in \F^k $ will be fetched for efficient computation and a high convergence rate.  
Finally, the ML training computes a prediction model $ \mlparam' $ from a set of feature vectors $ \{\vect{x}'_i\} $ as well as their labels $ \{y_i\} $, while ML inference computes the label $ y $ from the feature vector $ \vect{x}' $ of a new observation using the prediction model $ \mlparam' $.

In this paper, 
we focus on the ML inference pipeline (MLIP), in which the client collects raw data, and the server processes the data in multiple stages (i.e., preprocessing, feature extraction, ML classification) to obtain the final inference result.
At each stage, the server can employ its private ML model parameters obtained from its training pipeline to process the client data.
We denote such MLIP functionality as \revnew{$ y \gets \funcml (\mlparam, \vect{x}) $}, where $ \vect{x} \in \F^m $ is the data sample, \revnew{$ \mlparam \in \F^{n} $} is MLIP model parameters in all stages, and $ y \in \F $ is the inference result.

\section{Models}\label{sec:models}

\PP{System and threat models} Our system consists of two parties, including the client and the server.
The server holds well-trained MLIP model parameters $ \mlparam $
and provides an interface for the client to classify her data sample $ \vect{x} $ using its model $ \mlparam $.

We consider the client and server to mutually distrust each other. 
The adversarial server can be malicious, in which it may process the client's query arbitrarily.
On the other hand, {the client is semi-honest, in which she is curious about the server's model parameters.}
In this setting, we aim to achieve inference integrity and model privacy.
To enable inference integrity, the server first commits to its model $ \mlparam $.
Given a client request, the server computes the inference result $ y $ along with a proof $ \pi $ to convince the client that the result is indeed computed from the committed model rather than an arbitrary answer.
To ensure model privacy, the proof $ \pi $ should not leak any information about the model $ \mlparam $.

Formally speaking, a zero-knowledge MLIP is a tuple of algorithms $\zkmlip = (\Gen,\Commit,\Infer,\Verify)$ as follows
\begin{itemize}
\item $\pp \gets  \zkmlip.\Gen(1^\secparam,n)$: Given a security parameter $ \secparam $ and a bound on the size of the MLIP model parameters $ n $, it outputs public parameters $ \pp $. 
\item $ \cm \gets \zkmlip.\Commit(\mlparam,r,\pp)$: Given MLIP parameters $ \mlparam $, it outputs a commitment $ \cm $ under randomness $ r $. 
\item $(y,\pi) \gets \zkmlip.\Infer(\mlparam,\vect{x},\pp)$: Given MLIP model parameters $ \mlparam $ and a data sample $ \vect{x} $, it outputs the inference result $ y  = \funcml(\mlparam,\vect{x})$ and the  proof  $ \pi $. 
\item $\{0, 1\} \gets \zkmlip.\Verify(\cm,\vect{x},y,\pi,\pp)$: Given a commitment $ \cm $, a sample $ \vect{x} $, an inference result $ y $, and a proof $ \pi $, 
it outputs $ 1 $ if $ \pi $ is the valid proof for $ y =\funcml(\mlparam,\vect{x})$ and {$\cm = \Commit(\mlparam, r, \pp)$}; otherwise it outputs 0. 
%

\end{itemize}

\PP{Security model} We define the security definition of zero-knowledge MLIP that captures inference integrity and model privacy in the integrity proof as follows.

\begin{Definition}[zero-knowledge MLIP] \label{def:sec:zkmlip}
	A scheme is zero-knowledge MLIP if it satisfies the following properties.
	\begin{itemize}
		\item \textbf{Completeness}.  For any  {$ \mlparam \in \F^{n} $} and $ \vect{x} \in \F^m$,  
		$ \pp \gets  \zkmlip.\Gen(1^\secparam,n)$, $\cm \gets \zkmlip.\Commit(\mlparam,r,\pp)$, $(y,\pi) \gets \zkmlip.\Infer(\mlparam,\vect{x},\pp)$, it holds that
	\begin{gather*}
	\Pr \left[
		 \zkmlip.\Verify(\cm,\vect{x},y,\pi,\pp) = 1
	\right] = 1
	\end{gather*}
		\item \textbf{Soundness}. For any $\mathrm{PPT}$ adversary $\mathcal{A}$, it holds that
		
				\begin{gather*}
				\Pr \left[
				\begin{aligned}
					\pp \gets  \zkmlip.\Gen(1^\secparam,n) \\
					(\cm^*,\mlparam^*,\vect{x},y^*,\pi^*,r) \gets \Adv(\pp) \\
					\cm^* = \zkmlip.\Commit(\mlparam^*,r,\pp)\\
					\zkmlip.\Verify(\cm^*,\vect{x},y^*,\pi^*,\pp) = 1  \\
					\funcml(\mlparam^*,\vect{x}) \ne y^* 			
				\end{aligned}
				\right] \leq \mathsf{negl}(\secparam)
			\end{gather*}
		
		\item \textbf{Zero-knowledge.} For any MLIP model $ \mlparam \in \F^n $ and PPT algorithm $ \Adv $, there exists simulator $ \Sim = (\Sim_1, \Sim_2) $ such that
\begin{gather*}
		\begin{aligned}
		\Pr \left[
		\begin{array}{l}
			\Adv(\cm,\vect{x},y,\pi,\pp)= 1 
		\end{array}
		\middle \vert
		\begin{array}{r}
				 \pp \gets \zkmlip.\Gen(1^\secparam,n)\\
				\cm \gets \zkmlip.\Commit(\mlparam,r,\pp) \\
				\vect{x} \gets \Adv(\cm,\pp) \\
				(y,\pi) \gets \zkmlip.\Infer(\mlparam, \vect{x}, \pp)
		\end{array}
		\right] \stackrel{c}{\approx}
		\Pr \left[
		\begin{array}{l}
			\Adv(\cm,\vect{x},y,\pi,\pp) = 1
		\end{array} 
		\middle \vert 
		\begin{array}{r}
			(\cm,\pp) \gets \Sim_1(1^\secparam,n,r) \\
			\vect{x} \gets \Adv(\cm,\pp) \\
		    (y,\pi) \gets  \Sim_2^{\Adv}(\cm, \vect{x}, r, \pp), \text{given} \\ \text{oracle access to } y = \funcml(\mlparam,\vect{x}) 
			\\
		\end{array}
		\right]
		\end{aligned}
\end{gather*}	
	\end{itemize}
\end{Definition}

\PP{Out-of-scope attacks}
Our security definition captures the inference integrity and the model privacy in the integrity proof $ \pi $.
There exist model stealing attacks \cite{tramer2016stealing,chandrasekaran2020exploring} that target only the inference result $ y $ to reconstruct the model $ \mat{w} $.
In this paper, we do not focus on addressing such vulnerabilities.
It is because there exist independent studies that address these vulnerabilities (e.g., \cite{272262,tramer2016stealing,chandrasekaran2020exploring,kesarwani2018model,lee2018defending}) and, with some efforts, they can be integrated orthogonally into our scheme to protect $ \mat{w} $ from both $ y $ and $ \pi $.
For example, by simply limiting the inference result information (i.e., return only the predicted label like our scheme currently offers), it makes the attack become $ 50$-$100 \times$ more difficult \cite{tramer2016stealing}.  
We elaborate all these approaches in \autoref{sec:counterMEA}.
Our main goal is to ensure $ \mat{w} $ is not leaked from $ \pi $ via zero-knowledge
so that the leakage from $ y $ can be sealed or mitigated independently by these techniques.
For curious readers, we also show how $\pi$ may leak significant information about $ \mat{w} $ if it is not zero-knowledge in \autoref{sec:ModelLeakageFromNonZKPoI}.

We also do not consider model poisoning/backdoor attacks (e.g., \cite{ML:attack:backdoor:salem2020don,ML:attack:backdoor:salem2020dynamic}), in which the adversarial server may target adversarial behaviors on certain data samples while maintaining an overall high level of accuracy.
Mitigating such attacks requires analyzing the model parameters (e.g., \cite{backdoor:defense:model-based:liu2019abs}, which may be highly challenging in our setting, where the model privacy is preserved.
Thus, we leave this threat model as an open research problem for future investigation.

\vspace{-.7em}
\section{Our Proposed Zero-Knowledge MLIP Framework} \label{sec:proposedmethod}
In this section, we present the detailed construction of our framework.
We start by giving an overview.

\PP{Overview} Our \sys~framework contains three processing phases, 
including data (pre)processing, feature extraction, and ML classification, as shown in \autoref{fig:generalML}.  
We adopt ML algorithms for each phase
including Discrete Wavelet Transformation (DWT) \cite{vonesch2007generalized} for data preprocessing,
Principal Components Analysis (PCA) \cite{wold1987principal} for feature extraction,
and Support Vector Machine (SVM) \cite{2003Radius} for classification.
We focus on these algorithms because they were well-established 
in various systems and applications with high efficiency \cite{SVM:PCA:DWT:martis2013ecg,SVM:PCA:DWT:luz2016ecg}. 
\sys~permits to verify a data sample was computed correctly 
with DWT, PCA, and SVM without leaking the parameters at each phase including, for example, low-pass and high-pass filters in DWT; 
mean vector and eigenvectors in PCA; and support vectors in SVM.

In \sys, the server first commits to the model parameters of each ML algorithm and provides an interface for the client to process her data sample based on the committed parameters.
To demonstrate the validity of the committed model, the server can publish a zero-knowledge Proof-of-Accuracy (zkPoA) 
to demonstrate that the committed model maintains a desirable accuracy on public datasets with ground truth labels. 
zkPoA permits the client to attest to the genuineness and the effectiveness of the server's committed model 
before using the inference service on her data sample.
zkPoA can be derived from zero-knowledge proof of inference of individual samples.
We show how to construct zkPoA for our scheme in \autoref{sec:zkPoA}.

In the following sections, 
we first present new gadgets for critical ML operations (e.g., max/min, absolute).
Notice that our proposed gadgets are not limited to the ML algorithms selected above.
They can be used to prove other useful ML kernels (\autoref{sec:other_kernel}) and deep learning components (\autoref{sec:proveDL}).
%
We then present our techniques for proving DWT, PCA, and SVM more efficiently than the generic approaches.
Finally, we show how to construct a zkPoA scheme to attest to the effectiveness of the committed model on public datasets. 

\subsection{Gadgets}
\label{gadgets and functions}
A gadget is an intermediate constraint system consisting of a set of arithmetic constraints for proving a particular statement in the higher-level protocols. 

\subsubsection{Building Blocks}We first present building block gadgets that were previously proposed.

\PP{Permutation gadget \cite{zhang2020zero}}
Given two vectors $ \vect{v}, \vect{v}' \in \F^n $, $\mathsf{Perm}(\vect{v},\vect{v}')$ permits to prove that  $ \vect{v} $ is the permutation of $ \vect{v}' $, i.e., $\vect{v}[i] = \vect{v}'[{\sigma(i)}] $ for $i \in [1,n]$ according to some permutation $ \sigma$.
This can be done by showing that their characteristic polynomial evaluates to the same value at a random point $ \alpha $ chosen by the verifier as

	\begin{equation*}
		\prod_{i=1}^{n} (\vect{v}[i]-\alpha) = \prod_{i=1}^{n} (\vect{v}'[i]-\alpha)
	\end{equation*}

Due to Schwartz-Zippel Lemma \cite{schwartz1980fast}, the soundness error of the permutation test is $\frac{n}{|\mathbb{F}|} = \mathsf{negl}(\secparam)$.

\PP{Binarization  gadget \cite{sasson2014zerocash}} 
Given a vector $ \vect{v} \in \F^n$ and a value $ a \in \F $,
binarization gadget $\mathsf{Bin}(a,\vect{v}, n)$ permits to prove that $ \vect{v} $ is a binary representation of $ a $.
This can be done by showing that 

	\begin{equation*}
		\begin{cases}
			\vect{v}[i] \times \vect{v}[i] = \vect{v}[i] \text{ for } i \in [1,n]\\
			\sum_{i=1}^{n}\vect{v}[i] \cdot  2^{i-1} = a
		\end{cases} 
	\end{equation*}

\subsubsection{New Gadgets for Zero-Knowledge MLIP}
\label{sec:function}

We now construct new gadgets that are needed in our $ \sys $ scheme. These gadgets can be used to prove other ML algorithms that incur the same operations.

\PP{Exponent gadget} 
Given two values $ b, x \in \F $, we propose a gadget $ \mathsf{Exp}(b,a,x) $ to prove $ b = a^x $ for public value $ a \in \F $\footnote{The exponent gadget was briefly mentioned in \cite{zhang2020zero}, but no concrete constraints were given. We give concrete arithmetic constraints for proving exponent in arithmetic circuits.}.
This can be done using the multiplication tree and the binarization gadget ($\mathsf{Bin}$).
Let $ \vect{v} \in \F^n$ be an auxiliary witness. It suffices to show that

\begin{equation*}\label{gadget:exp}
		\begin{cases}
			\mathsf{Bin}(x,\vect{v}, n) \\
			b = \prod_{i=1}^{n} (a^{2^{i-1}} \cdot  \vect{v}[i] + (1 - \vect{v}[i]))\\
		\end{cases} 
\end{equation*}

\noindent\textbf{GreaterThan  gadget}. 
Given two values $ a,b \in \F $, we create a gadget $\mathsf{GT}(a,b)$ to prove that 
$ a > b $.
The main idea is to compute an auxiliary witness $ c := 2^n + (a-b) $, where $ n $ is the length of the binary representation of $ a $ and $ b $, and show that the most significant bit of $ c $ is equal to 1. 
Let $ \vect{c}  \in \F^{n+1}$ and $\vect{a}, \vect{b} \in \F^{n}$ be additional auxiliary witnesses.
The set of arithmetic constraints to prove $ a > b $ is 

	\begin{equation*}
		\begin{cases}
			c = 2^n + a - b  \\
			\mathsf{Bin}(a,\vect{a},n)\\
			\mathsf{Bin}(b,\vect{b},n)\\
			\mathsf{Bin}(c,\vect{c},n+1)\\
			\vect{c}[n+1] = 1\\
		\end{cases} 
	\end{equation*}

\PP{Maximum/Minimum gadget}
Given a value $ v \in \F $ and an array $\vect{a} \in \F^n$,
we create a gadget $\mathsf{Max}(v,\vect{a})$ (resp. $\mathsf{Min}(v,\vect{a})$) to prove that $ v $ is the maximum (resp. minimum) value in $\vect{a}$.
The idea is to harness $\mathsf{Perm}$ and $\mathsf{GT}$ gadgets to prove that $ v $ is equal to the first element of the permuted array of $ \vect{a} $, whose first element is the largest (resp. minimum) value. %
Specifically, 
to prove $v = \max(\vect{a})$, it suffices
to show $(\romannumeral1)$ $ v = \vect{a}'[1] $, ($\romannumeral2$) $\vect{a}'[1] > \vect{a}'[i]$ for all $i \in [2, n]$, and ($\romannumeral3$) $\vect{a}'$ is the permutation of $\vect{a}$.
Let $ \vect{a}' \in \F^n$ be an auxiliary witness. 
The set of arithmetic constraints to prove a maximum value in an array is 

	\begin{equation*}
		\begin{cases}
			\mathsf{GT}(\vect{a}'[1],\vect{a}'[i]) \text{ for all } i \in [2,n]\\
			v = \vect{a}'[1]\\
			\mathsf{Perm}(\vect{a},\vect{a}')
		\end{cases} 
	\end{equation*}

The constraints to prove a minimum value in an array can be defined analogously.

\begin{table}[t!]
    \centering
    \small
    \caption{Notation table.}
    \begin{tabular}{|lll|}
    \hline
        \multicolumn{2}{|l}{\textbf{Variables}} &  \textbf{Description}\\\hline
        \multicolumn{3}{|l|}{\textit{DWT components}}\\
        & $\vect{x} \in \F^m$ & Sample input of size $m$ to DWT\\ 
        & $\vect{h}, \vect{\bar{\vect{h}}}\in \F^c$ & low-pass filter of size $c$ and its inverse \\ 
        & $\vect{g}, \bar{\vect{g}}\in \F^c$ & high-pass filter of size $c$ and its inverse\\ 
        
        &$\eta$ & Filter threshold\\
        \hline
        \multicolumn{3}{|l|}{\textit{PCA components}}\\
        
        & $\hat{\vect{x}} \in \F^m$ & Sample input of size $m$ to PCA \\ 
        
        & $\bar{\vect{x}} \in \F^m$ & Mean vector  \\ 
        
        & $\vect{V} = [\vect{v}_1^T, ...,\vect{v}_m^T]$ & Eigenvectors \\ 
        
        & $(\lambda_1,...,\lambda_m)$ & Eigenvalues \\ 
        
        & $ k $ & Size of PCA output \\ \hline
        
        \multicolumn{3}{|l|}{\textit{SVM components}}\\
        
        & $ \phi $ & kernel function \\ 
        
        & $ \gamma $ & RBF kernel parameter \\ 
        
        & $ \vect{x}_i^{(\hat{c})}$ & Support vectors for class $\hat{c}$ \\ 
        
        & $ \vect{w}^{(\hat{c})}, b^{(\hat{c})}$ & Weights and bias for class $\hat{c}$ \\ 
        
        & $ y_i^{(\hat{c})} \in \{0,1\}$ & Label of class $ \hat{c} $ \\ 
        & $ \delta^{(\hat{c})}$ & Coefficients of class $\hat{c}$ in RBF kernel \\ 
        
        & $f^{(\hat{c})}$ & Decision function of class $\hat{c}$\\ 
        \hline
        \multicolumn{3}{|l|}{\textit{Proof components}}\\
        & $\sigma$ & Permutation function \\ 
        & $\lambda$ & Security parameter \\ 
        & $\pi$ & Proof \\ 
        & $\mlparam$ & Witness \\
        & $\mathsf{aux}$ & Auxiliary witness \\
        & $ \cm $ & Commitment \\ 
        & $\alpha, \bar{\alpha}, \beta$ & Random challenges \\ \hline
    \end{tabular}

    \label{tab:notation}\vspace{-.7em}
\end{table}

\PP{Absolute gadget}
Given $ a',a \in \F $,
we create gadget $\mathsf{Abs}(a',a)$ to prove that $a'$ is the absolute value of $ a $, i.e., $a = a'$ or $- a = a'$.
The  idea is to compute  $c = a + 2^n$, where $ n $ is the length of the binary representation of $ a $, and show that the most significant bit of $ c $ represents the sign difference of  $ a $ and $a'$. Let $\vect{c} \in \F^{n+1}$ and $\vect{a} \in \F^n$ be auxiliary witnesses, 
the set of arithmetic constraints to show that $ a' $ is the absolute value of $a $ is %

\vspace{-.7em}
	\begin{equation*}
		\begin{cases}
			c = a + 2^n \\
			\mathsf{Bin}(a,\vect{a},n)\\
			\mathsf{Bin}(c,\vect{c},n+1)\\
			(1-\vect{c}[n+1])(a+a')+\vect{c}[n+1](a-a')=0\\
		\end{cases} 
	\end{equation*}

\subsection{Our Proposed Scheme}
\label{sec:ml:ml ins}

We now give the detailed construction of our \sys~scheme with DWT, PCA, and SVM algorithms. We provide the overview of each algorithm and show how to prove it with a small number of constraints. We summarize all the variables and notation being used for our detailed description in \autoref{tab:notation}.

\vspace{-.7em}
\subsubsection{DWT-Based Data Preprocessing}
\label{sec:subsubsec:dwt}

DWT \cite{vonesch2007generalized} exerts the wavelet coefficients on the raw data sample to project it to the wavelet domain for efficient preprocessing.  
A DWT algorithm contains three main operations, including decomposition, thresholding, and reconstruction.
The decomposition transforms the raw input from the spatial/time domain to the wavelet domain consisting of approximation and detail coefficients.
The thresholding is then applied to filter some detail coefficients,
which generally contain noise. 
Finally, the reconstruction is applied to reconstruct the original data after noise reduction.
Such decomposition and thresholding processes can be applied recursively until a small constant number of coefficients is obtained. 
Let $\vect{x}\in \mathbb{F}^m$ be the input data sample of length $ m $, $t_\ell := \frac{m}{2^\ell}$, $t'_\ell := \frac{m}{2^{\ell-1}}$. 
The DWT computes the frequency component $ \vect{z}_\ell \in \mathbb{F}^{t'_{\ell}}$ at the recursion level $ \ell \ge 1 $ as

	\begin{equation}\label{eq:dwt:decompose}
		\begin{aligned}
			\vect{z}_{\ell}[i] & =  \sum_{j=1}^{c} \vect{h}[j] \cdot \vect{z}_{\ell-1}[(2i+j-2)_{\text{mod }t'_\ell}] \\
			\vect{z}_{\ell}[i+t_{\ell}] &= \sum_{j=1}^{c} \vect{g}[j] \cdot \vect{z}_{\ell-1}[(2i+j-2)_{\text{mod }t'_{\ell}}] 
		\end{aligned}
	\end{equation}

for $i \in [1,t_{\ell}]$, where $ \vect{h}, \vect{g}\in \mathbb{F}^c$ are low-pass and high-pass filters respectively, and $\vect{z}_0 = \vect{x}$. 
The thresholding is applied to compute high-frequency components (i.e., detail coefficients) as

	\begin{gather}\label{eq:dwt:threshold}
		\begin{aligned}
			\vect{z}'_{\ell}[i] & = \vect{z}_{\ell}[i]\\
			\vect{z}'_{\ell}[{i+t_\ell}] & = \left\{
			\begin{aligned}
				& \mathsf{sign}(\vect{z}_{\ell}[{i+t_\ell}]) (\vect{z}_{\ell}[{i+t_\ell}] - \eta) & \text{ if } \left|\vect{z}_{\ell}[{i+t_\ell}]\right| - \eta > 0 \\ 
				& 0  & \text{ if } \left|\vect{z}_{\ell}[{i+t_\ell}]\right| - \eta < 0
			\end{aligned}		
			\right\}
		\end{aligned}
	\end{gather}
%
for $i \in [1,t_{\ell}]$, where $ \eta $ is the public threshold parameter, $\mathsf{sign}(x)$ returns the sign of $ x $ (i.e., $1$ if $x \ge 0$, and $-1$ otherwise). 
The decomposition and thresholding can be applied recursively until $t_{\ell} < c$, or the number of rounds reaches a set value.
Finally, the reconstructed data $ \vect{\hat{x}}_\ell \in \mathbb{F}^{t'_\ell}$ at recursion level $ \ell $ is computed as

	\begin{gather}\label{eq:dwt:reconstruct}
		\begin{aligned}
			\vect{\hat{x}}_\ell[2i-1] = \sum\limits_{j=1}^{c/2}(\vect{\bar{h}}&[2j-1]\cdot \vect{z}'_\ell[(i+j-1)_{\text{mod }t_\ell}]+\vect{\bar{h}}[2j]\cdot \vect{z}'_\ell[t_\ell+(i+j-1)_{\text{mod }t_\ell}]) \\
			\vect{\hat{x}}_\ell[2i] = \sum\limits_{j=1}^{c/2}(\vect{\bar{g}}&[2j-1]\cdot \vect{z}'_\ell[(i+j-1)_{\text{mod }t_\ell}]+\vect{\bar{g}}[2j]\cdot \vect{z}'_\ell[t_\ell+(i+j-1)_{\text{mod }t_\ell}])
		\end{aligned}
	\end{gather}
%
for $ i \in [1, t_{\ell}]$, $\vect{\bar{h}}, \vect{\bar{g}} \in \F^c$ are the coefficients of the inverse low-pass and high-pass filters, respectively. In summary, the DWT model parameters are $\vect{h}, \vect{g}, \vect{\bar{h}}, \vect{\bar{g}}, \eta$. The size of the model parameter is $ 4c+1$, where $c$ depends on the concrete DWT algorithm used in practice, e.g., $ c=4 $ in DB-4 algorithm. 

\PP{Proving DWT computation} 
We can see that \autoref{eq:dwt:decompose} incurs $8m(1-\frac{1}{2^l})$ constraints, where $ m $ is the length of the data sample, $l$ is the number of recursion levels.
We propose a novel method to prove DWT computation in a more efficient way
using our proposed split technique along with the product of sums and random linear combination.
Our optimization reduces the complexity of proving the decomposition and reconstruction from $ O(m)$ to $ O(\log m)$. Furthermore, if the recursion level $ l $ is set to a constant, the complexity can be reduced to $ O(1) $.
Specifically, we first split each element in $\vect{z}_\ell \in \mathbb{F}^{t'_{\ell}}$ into two parts as
	\begin{equation}\label{eq:dwt:decompose:split}
		\begin{aligned}
			\vect{z}_\ell[i]^{(\tt 1)}  & =  \sum_{k=1}^{c/2}\vect{h}[2k-1]\cdot \vect{z}_{\ell-1}[(2k+2i-3)_{\text{mod }t'_{\ell}}]\\
			\vect{z}_\ell[i]^{(\tt 2)}  & =  \sum_{k=1}^{c/2}\vect{h}[2k] \cdot \vect{z}_{\ell-1}[(2k+2i-2)_{\text{mod }t'_{\ell}}]\\
			\vect{z}_\ell[i+t_{\ell}]^{\tt (1)} & =  \sum_{k=1}^{c/2}\vect{g}[2k-1]\cdot \vect{z}_{\ell-1}[(2k+2i-3)_{\text{mod }t'_{\ell}}]\\
			\vect{z}_\ell[i+t_{\ell}]^{\tt (2)} & =  \sum_{k=1}^{c/2}\vect{g}[2k] \cdot \vect{z}_{\ell-1}[(2k+2i-2)_{\text{mod }t'_{\ell}}]\\
		\end{aligned}
	\end{equation}
for $i \in [1, t_{\ell}]$. Let $ \alpha \in \mathbb{F} $ be a random {scalar} chosen by the verifier, the prover can prove \autoref{eq:dwt:decompose:split} holds such that

\begin{equation}\label{eq:dwt:decompose:split:randlin}
	\begin{aligned}
		&\sum_{i=1}^{t_{\ell}} \alpha^i  \vect{z}_\ell[i]^{(\tt 1)}   =  \sum_{i=1}^{t_{\ell}} \alpha^i \cdot \sum_{k=1}^{c/2}\vect{h}[2k-1]\cdot \vect{z}_{\ell-1}[(2k+2i-3)_{\text{mod }t'_{\ell}}] \\
		&\sum_{i=1}^{t_{\ell}} \alpha^i \vect{z}_\ell[i]^{(\tt2)}   =  \sum_{i=1}^{t_{\ell}} \alpha^i \cdot \sum_{k=1}^{c/2}\vect{h}[2k] \cdot \vect{z}_{\ell-1}[(2k+2i-2)_{\text{mod }t'_{\ell}}] \\
		&\sum_{i=1}^{t_{\ell}} \alpha^i  \vect{z}_\ell[i+t_{\ell}]^{(\tt 1)}  = \sum_{i=1}^{t_{\ell}} \alpha^i  \cdot \sum_{k=1}^{c/2}\vect{g}[2k-1]\cdot \vect{z}_{\ell-1}[(2k+2i-3)_{\text{mod }t'_{\ell}}]\\
		&\sum_{i=1}^{t_{\ell}} \alpha^i  \vect{z}_\ell[i+t_{\ell}]^{(\tt 2)}  = \sum_{i=1}^{t_{\ell}} \alpha^i  \cdot \sum_{k=1}^{c/2}\vect{g}[2k] \cdot \vect{z}_{\ell-1}[(2k+2i-2)_{\text{mod }t'_{\ell}}]\\
	\end{aligned}
\end{equation}

We convert \autoref{eq:dwt:decompose:split:randlin} to the product of sums as
{\begin{gather}
	\begin{aligned} \label{eq:dwt:decompose:final}
		\begin{split}
			\sum\limits_{i=1}^{t_{\ell}}\alpha^{\frac{c}{2}+i-2}\vect{z}_\ell[i] & =  
			\sum_{k=1}^{c/2}\alpha^{\frac{c}{2}-k}  \vect{h}[2k-1] \cdot \sum\limits_{i=1}^{t_{\ell}} \alpha^{i-1} \vect{z}_{\ell-1}[2i-1] + \sum_{k=1} ^{c/2}\alpha^{\frac{c}{2}-k} \cdot \vect{h}[2k]\cdot \sum\limits_{i=1}^{t_{\ell}} \alpha^{i-1} \vect{z}_{\ell-1}[2i] + (\alpha^{t_{\ell}}-1)\sum_{q=1}^{\frac{c}{2}-1}\alpha^{q-1} \\
			& \quad \cdot \sum_{p=1}^{q} \left(\vect{z}_{\ell-1}[2p]\vect{h}[c-2q+2p] + \vect{z}_{\ell-1}[2p-1]\vect{h}[c-2q+2p-1] \right)\\
			\sum\limits_{i=1}^{t_{\ell}}\alpha^{\frac{c}{2}+i-2}\vect{z}_\ell[i+t_{\ell}] & =  
			\sum_{k=1}^{c/2}\alpha^{\frac{c}{2}-k}  \vect{g}[2k-1]\cdot \sum\limits_{i=1}^{t_{\ell}} \alpha^{i-1} \vect{z}_{\ell-1}[2i-1] + \sum_{k=1} ^{c/2}\alpha^{\frac{c}{2}-k}  \vect{g}[2k]\cdot \sum\limits_{i=1}^{t_{\ell}} \alpha^{i-1} \vect{z}_{\ell-1}[2i] + (\alpha^{t_{\ell}}-1)\sum_{q=1}^{\frac{c}{2}-1}\alpha^{q-1} \\
			& \quad \cdot \sum_{p=1}^{q} \left(\vect{z}_{\ell-1}[2p]\vect{g}[c-2q+2p]+\vect{z}_{\ell-1}[2p-1]\vect{g}[c-2q+2p-1]\right)\\
		\end{split}
	\end{aligned}
\end{gather}}

In \autoref{eq:dwt:decompose:final}, the number of constraints for proving DWT decomposition is reduced from $ mc $ to  $c(\frac{c}{2}-1)+4$.
To aid understanding, we present a toy example of our split technique in \autoref{sec:example:split}.
To prove the thresholding computation in \autoref{eq:dwt:threshold},
we employ the $\mathsf{GT}$ gadget, such that for $i \in [1, t_\ell]$:

\vspace{-.7em}
\begin{equation}\label{eq:thresholding}
	\begin{cases}
		\mathsf{GT}(\vect{z}_\ell[i+t_{\ell}], \eta) \text{ for all } \vect{z}'_\ell[i+t_{\ell}] \neq 0\\
		\mathsf{GT}(\eta, \vect{z}_\ell[i+t_{\ell}]) \text{ for all } \vect{z}'_\ell[i+t_{\ell}] = 0\\
		\vect{z}'_{\ell}[i] - \vect{z}_\ell [i] = 0 
	\end{cases} 
\end{equation}

In our protocol, the prover provides $|\vect{z}_\ell[i]|$ and $\mathsf{sign}(\vect{z}_\ell[i])$ as the auxiliary witnesses so that the number of constraints reduces from $5n+14$ to $3n+9$ for each $\vect{z}_\ell[i+t_{\ell}]$, where $n$ is the length of the binary representation of $\vect{z}_\ell[i+t_{\ell}]$. 

The final step is proving the DWT reconstruction, which is analog to proving the decomposition. Let $\bar{\alpha} \in \mathbb{F}$ be a random challenge chosen by the verifier. The prover can prove DWT reconstruction in \autoref{eq:dwt:reconstruct} such that

{\begin{gather}
	\begin{aligned} \label{eq:dwt:reconstruct:prove:final}
		\begin{split}
			\sum\limits_{k=1}^{t_{\ell}}\bar{\alpha}^{\frac{c}{2}+i-2} \vect{\hat{x}}_\ell[(2k+1)_{\text{mod } t'_\ell}]  &= \sum_{k=1}^{c/2}\bar{\alpha}^{\frac{c}{2}-k}  \vect{\bar{h}}[2k-1]\cdot \sum\limits_{i=1}^{t_{\ell}} \bar{\alpha}^{i-1} \vect{z}'_{\ell-1}[i]  +  \sum_{k=1} ^{c/2}\bar{\alpha}^{\frac{c}{2}-k}  \vect{\bar{h}}[2k] \cdot \sum\limits_{i=1}^{t_{\ell}} \bar{\alpha}^{i-1} \vect{z}'_{\ell-1}[i+t_{\ell}] \\
			& +(\bar{\alpha}^{t_{\ell}}-1)\cdot \sum_{q=1}^{\frac{c}{2}-1}\bar{\alpha}^{q-1}\cdot\sum_{p=1}^{q}\left(\vect{z}'_{\ell-1}[p]\vect{\bar{h}}[c-2q+2p]+\vect{z}'_{\ell-1}[p+t_{\ell}]\vect{\bar{h}}[c-2q+2p-1]\right)\\
			\sum\limits_{k=1}^{t_{\ell}}\bar{\alpha}^{\frac{c}{2}+i-2} \vect{\hat{x}}_\ell[(2k)_{\text{mod }t'_{\ell}}]  &= \sum_{k=1}^{c/2}\bar{\alpha}^{\frac{c}{2}-k} \vect{\bar{g}}[2k-1]\cdot \sum\limits_{i=1}^{t_{\ell}} \bar{\alpha}^{i-1} \vect{z}'_{\ell-1}[i] +\sum_{k=1} ^{c/2}\bar{\alpha}^{\frac{c}{2}-k}  \vect{\bar{g}}[2k] \cdot \sum\limits_{i=1}^{t_{\ell}}\bar{\alpha}^{i-1} \vect{z}'_{\ell-1}[i+t_{\ell}] \\
			& +(\bar{\alpha}^{t_{\ell}}-1)\sum_{q=1}^{\frac{c}{2}-1}\bar{\alpha}^{q-1}\cdot \sum_{p=1}^{q}\left(\vect{z}'_{\ell-1}[p]\vect{\bar{g}}[c-2q+2p]+\vect{z}’_{\ell-1}[p+t_{\ell}]\vect{\bar{g}}[c-2q+2p-1]\right)\\
		\end{split}
	\end{aligned}
\end{gather}}

\subsubsection{PCA-Based Feature Extraction}
PCA \cite{wold1987principal} is a method to reduce the dimensionality of the data input by representing the most significant characteristics of $ \vect{\hat{x}} \in \mathbb{F}^m $ in a smaller feature vector with minimal information loss (i.e., eigenvalues). The PCA training computes a mean vector $\vect{\bar{x}} \in \mathbb{F}^m$ for all data samples $\{\vect{\hat{x}}_i\}_{i=1}^N$ as $\bar{\vect{x}} = \frac{\sum_i \vect{\hat{x}}_i}{N} $ {, where $N$ is the number of samples in the training set.}
A covariance matrix is then computed as 
%
$		\mat{S} = \frac{1}{N}\sum_{i=1}^{N} (\vect{\hat{x}}_i - \vect{\bar{x}}) (\vect{\hat{x}}_i - \vect{\bar{x}})^\top$.
%
The PCA training aims at finding eigenvectors $\mat{V}=[\vect{v}_1^\top,\dots,\vect{v}_m^\top]$ and eigenvalues $(\lambda_1,\dots,\lambda_m)$ of $ \mat{S} $ such that
%
$		\mat{S} \times \mat{V} =  \mat{V} \times \mat{\Lambda}$
%
\noindent where $ \mat{\Lambda} = \mathsf{diag}(\lambda_1,\dots,\lambda_m) $. 
To reduce the dimension while retaining the most information about data distribution, 
we select $ k $ eigenvectors $\mat{V}'=[\vect{v}_{i_1}^\top,\dots,\vect{v}_{i_k}^\top]$ corresponding with $ k $ largest eigenvalues $(\lambda_{i_1},\dots,\lambda_{i_k})$. To this end, the server retains the eigenvectors $ \mat{V}' $ and the mean vector $\vect{\bar{x}}$ as  model parameters.
In the inference phase, given a new observation $\vect{\hat{x}}$, the feature vector of  $\vect{\hat{x}}$ can be computed via PCA as 
\begin{equation}\label{eq:PCA}
	\vect{\tilde{x}} = (\vect{\hat{x}} - \bar{\vect{x}}) \times \vect{V}'
\end{equation}

\PP{Proving PCA computation} 
There are $O(m \cdot k)$ constraints in \autoref{eq:PCA}, where $ m $ is the input {dimension and} $ k $ is the feature vector dimension. We reduce the number of constraints of proving PCA computation from $O(m \cdot k )$ to $O(m)$ using the random linear combination by using the powers of a random challenge chosen by the verifier. This transformation converts variables' multiplication to constant multiplication, where the latter comes for free in R1CS, therefore reducing the computing complexity. Specifically,
\autoref{eq:PCA} is equivalent to 
%
		\begin{equation}\label{eq:pca:pos}
		\begin{aligned}
			\vect{\tilde{x}}[1] & = (\vect{\hat{x}} - \vect{\bar{x}}) \times \mat{V}'[1]  \\
			& \dots \\
			\vect{\tilde{x}}[k] & = (\vect{\hat{x}} - \vect{\bar{x}}) \times \mat{V}'[k] \\
		\end{aligned}
	\end{equation}
%
where $\mat{V}'[k]$ is the $k$th term in $ \mat{V}' $, e.g., {$\mat{V}'[k] = \vect{v}_{ik}^\top$}. Let $ \alpha \in \mathbb{F}$ be a random challenge chosen by the verifier. We apply the random linear combination to combine constraints in \autoref{eq:pca:pos}. Specifically, the prover can prove \autoref{eq:pca:pos} holds by proving that
%
		\begin{equation}\label{eq:pca:finalconstraints}
		\begin{aligned} 
			\sum_{i=1}^k \alpha^i \vect{\tilde{x}}[i] & =  (\vect{\hat{x}} - \vect{\bar{x}}) \times \sum_{i=1}^{k} \alpha^i \mat{V}'[i] \\
			& = \sum_{j=1}^m \left(\sum_{i=1}^k \alpha^i \mat{V'}[i]\right) \cdot (\vect{\hat{x}}[j] - \vect{\bar{x}}[j])	\end{aligned}
	\end{equation}

\noindent where $ \alpha^i $ is the power of the random challenge $\alpha$ computed by the prover, $ \mat{V}'$ is the eigenvector and $\bar{\vect{x}}$ is the mean vector.

\subsubsection{SVM Classification}
\label{sec:subsubsec:svm}
SVM \cite{2003Radius} is a supervised ML for classification problems by finding optimal hyperplane(s) that maximizes the separation of the data samples to their potential labels.
Suppose the number of samples in the training set is $ N $. Let $\vect{x}_1,\dots,\vect{x}_N \in \F^{k} $ be the feature vector of data samples and $ y_1,\dots,y_N \in \{1,\dots,s\} $ be its corresponding label.
To deal with data non-linearity,
kernel SVM projects $\vect{x}_i $ to a higher dimension using a mapping function $ \Phi : \F^m \rightarrow \F^{m'}  $, where $ m' > m $ and applies a kernel function $ \phi(\vect{x}_i,\vect{x}_j) = \Phi(\vect{x}_i) \cdot \Phi(\vect{x}_j) $ for training and classifying computation.
Radial Basis Function (RBF) \cite{2003Radius} $	\phi_{\mathsf{rbf}}(\vect{x}_i, \vect{x}_j) = e^{-\gamma \cdot ||\vect{x}_i-\vect{x}_j||^2}$ is the most popular SVM kernel due to its effectiveness.

SVM was initially designed for binary classification, but it can be extended to multiclass classification by breaking down the multiclass problem into multiple \emph{one-to-rest} binary classification problems.
For each class $ \hat{c} $, data samples are assigned to two classes, where $ y^{(\hat{c} )}_i = 1 $ if $ y_i  = \hat{c}  $, otherwise $y^{(\hat{c} )}_i = 0$.
%

			%



%
The trainable parameter of SVM is the tuple $(\vect{x}^{(\hat{c})}_i,\delta_i^{(\hat{c})}, b^{\hat{c} })$, where for class $\hat{c}$, $ \vect{x}^{(\hat{c})}_i $ is the support vector, $ \delta_i^{(\hat{c})} $ is the coefficient, and $b^{(\hat{c})}$ is the bias. The range of $ i $ depends on $|\set{I}^{(\hat{c} )}| := |\{i : \delta^{(\hat{c} )}_i > 0 \}|$, which equals to the number of the support vectors for class $ \hat{c} $. Note that $\delta^{(\hat{c} )}_i \leq 0 $ are dropped during the training. The tuple $(\vect{x}^{(\hat{c})}_i,\delta_i^{(\hat{c})}, b^{\hat{c} })$ acts as the secret of the prover, which will be committed to prove the computation.

Given a new observation $\vect{\tilde{x}}\in \mathbb{F}^k$,  its label $ y $ can be predicted as

\begin{equation}\label{eq:SVM:decision}
	y =  \argmax_{\hat{c} }  \sum_{i\in \set{I}^{(\hat{c} )}} \delta_i^{(\hat{c} )} y^{(\hat{c} )}_i \phi(\vect{\tilde{x}}, \vect{x}_i^{(\hat{c})})+ b^{(\hat{c})}
\end{equation}

\PP{Proving multi-class SVM classification with RBF kernel}
Suppose $f^{(\hat{c} )} = \sum_{i\in \set{I}^{(\hat{c} )}} \delta_i^{(\hat{c} )} y^{(\hat{c} )}_i  \phi(\vect{\tilde{x}}, \vect{x}^{(\hat{c})}_i) + b^{(\hat{c} )}$ is the decision function's evaluation for each class $\hat{c} \in [1, s]$. 
To prove the SVM classification in \autoref{eq:SVM:decision}, we harness $\mathsf{Exp}$ and $\mathsf{Max}$ gadgets in \autoref{sec:function} to prove the exponent in the RBF kernel projection, and the class output being the maximum value among all evaluations, respectively.
We adopt the representation in \cite{zhang2020zero} where $f^{(\hat{c} )}$ is expanded to a value-index pair, i.e., $\vect{f} := \{(f^{(1)}, 1), (f^{(2)}, 2), \dots, (f^{(s)}, s)\}$. 
Let $$\vect{\bar{f}} := \{(\bar{f}^{(1)}, \sigma(1)), (\bar{f}^{(2)}, \sigma(2)), \dots, (\bar{f}^{(s)}, \sigma(s))\} $$ 
be the permutation of $\vect{f}$, where $\sigma(\cdot)$ is the permutation function such that $\bar{f}^{(\hat{c})} = f^{(\sigma(\hat{c}))}$ and $\bar{f}^{(1)}$ is the maximum value in $\vect{f} $.
The prover provides $\vect{\bar{f}}$ as the auxiliary witness and shows that the output label $y = \sigma(1)$. 
Let $\beta$ be a random challenge from the verifier, the prover binds each value-index pair in $\vect{f}$ and $\vect{\bar{f}}$ to a single value as

\vspace{-.7em}
\begin{equation} \label{eq:svm:1}
	\begin{aligned}
		p^{(\hat{c})} & = f^{(\hat{c})} + \beta \cdot \hat{c}\\
		\bar{p}^{(\hat{c})} & = \bar{f}^{(\hat{c})} + \beta \cdot \sigma(\hat{c})
	\end{aligned}
\end{equation}
and invokes a permutation check using $\mathsf{Perm}$ gadget, where $\beta$ is a random number chosen by $\Verifier$.
Let $l_{i}^{(\hat{c})} \in \F$ for $i \in \set{I}^{(\hat{c} )}, \hat{c}  \in [1,s] $, $[\bar{f}^{(1)}, \dots, \bar{f}^{(s)}]$ be the auxiliary witness used in the gadget $\mathsf{Max}$. Suppose $y$ is the claimed output label and $f^{(y)}$ is the evaluation of the corresponding decision function. 
Let $\vect{p} = \{p^{(\hat{c})}\}$ and $\vect{\bar{p}} = \{\bar{p}^{(\hat{c})}\}$ be intermediate vectors, where 
$p^{(\hat{c})}$ and $\bar{p}^{(\hat{c})}$ are computed by \autoref{eq:svm:1}, respectively.
The set of arithmetic constraints to prove \autoref{eq:SVM:decision} is

	\begin{equation}\label{eq:svm:final}
		\begin{cases}
			k_{i}^{(\hat{c} )}=-\gamma ||\vect{\tilde{x}}-\vect{x}^{(\hat{c})}_{i}||^2 \text{ for } i \in \set{I}^{(\hat{c} )}, \hat{c}  \in [1,s]\\	
			f^{(\hat{c} )} = \sum_{i\in \set{I}^{(\hat{c} )}} \delta_i^{(\hat{c} )} y^{(\hat{c} )}_i l_i^{(\hat{c} )} + b^{(\hat{c} )} \text{ for } \hat{c}  \in [1,s]\\
			\revnew{\mathsf{Exp}(l_{i}^{(\hat{c} )}, e, k_{i}^{(\hat{c} )}) \text{ for } i \in \set{I}^{(\hat{c} )}, \hat{c}  \in [1,s] }\\
			\mathsf{Max}(f^{(y)},[f^{(1)},\dots,f^{(s)}])\\
			\mathsf{Perm}(\vect{p}, \vect{\bar{p}})\\ 
			f^{(y)} + \beta \cdot y = \bar{p}^{(1)}
		\end{cases}
	\end{equation}

\PP{Proving other SVM kernels}
%
Our techniques can be used to prove other SVM kernels such as the polynomial kernel, Sigmoid kernel, etc. The polynomial kernel $\phi_{\mathsf{ply}}(\vect{x}_i, \vect{x}_j) = (\gamma \vect{x}_i^T \vect{x}_j + a)^b$ can be easily proved via addition and multiplication gates, where $\gamma, a, b$ are parameters. Although it is relatively easy to prove, the polynomial kernel usually achieves a lower accuracy than the RBF kernel \cite{svmandpoly}. Due to the space constraint, we show how to prove other kernels in \autoref{sec:other_kernel}.

\subsubsection{Putting Everything Together}
We combine everything together and present the complete algorithmic description of our \sys~scheme in \autoref{pro:sys}. 
We describe the functionality (\autoref{alg:DPS}) that processes a data sample $ \vect{x} \in \F^m $  with DWT (\autoref{alg:MLIP:DWTPCASVM}, lines \ref{alg:DPS:DWT:start}-\ref{alg:DPS:DWT:end}), PCA (line \ref{alg:DPS:PCA}), and SVM {(lines \ref{alg:DPS:SVM:start}-\ref{alg:DPS:SVM:end})}, and returns an inference result $ y $.

\begin{figure}[!t]
	\small
\begin{mdframed}
	\begin{Protocol}[\sys]\label{pro:sys}
	Let $\lambda$ be the security parameter.
		\begin{itemize}
		\item \underline{$\pp \gets  \sys.\Gen(1^\secparam)$:} Output $\pp \gets \zkp.\Gen(1^\secparam)$
		\item \underline{$ \hat{\cm} \gets \sys.\Commit(\mlparam,r,\pp)$:} Let $ \mlparam = (\vect{h}, \vect{g}, \vect{\bar{h}}, \vect{\bar{g}}, \eta, \vect{\bar{x}}, \vect{V}', \{\vect{x}_i, \delta_i^{(\hat{c} )},b^{(\hat{c} )}\}_{i\in \set{I}^{(\hat{c} )}, \hat{c} \in [1,s]}, \gamma)$. 
		Compute $\hat{\cm} \gets \zkp.\Commit(\mlparam, r,\pp)$, where $ r $ is randomness chosen by the server.
		\item \underline{$(y,\pi) \gets \sys.\Infer(\mlparam,\vect{x},\pp)$:} 
		\begin{enumerate}
		\item The server executes \autoref{alg:DPS} to compute $ y \gets  \mathsf{DPS}(\mlparam,\vect{x})$, and commits to all the auxiliary witnesses $\mathsf{aux}$ in \autoref{eq:dwt:decompose:final}, \autoref{eq:thresholding} \autoref{eq:dwt:reconstruct:prove:final}, \autoref{eq:pca:finalconstraints}, \autoref{eq:svm:final} as $\cm' \gets \zkp.\Commit(\mathsf{aux},r',\pp)$ under randomness $r'$ chosen by the server. 
		\item	Upon receiving the randomness  $ \vec{\alpha} $ chosen by the client for checking the random linear combination and maximum value, 
		the server invokes backend ZKP protocol to get the proof as $ \pi \gets \zkp.\Prover((\mlparam, \mathsf{aux}), \vect{x}, y, \pp)$. 
		The server sends $(y,\pi)$ to the client.
		\end{enumerate}
		\item \underline{$b \gets \sys.\Verify(\cm,\vect{x},y,\pi,\pp)$:} Let $ \cm =(\hat{\cm}, \cm')$, the client invokes $b \gets \zkp.\Verifier(\cm,\vect{x},y,\pi,\pp)$ and outputs $ b $.
		\end{itemize}
	\end{Protocol}
\end{mdframed} \vspace{-0.7em}
	\caption{Our \sys~Protocol.} \label{fig:sys:protocol} \vspace{-1.5em}
\end{figure}

\subsubsection{Zero-Knowledge Proof of Accuracy} \label{sec:zkPoA}
We construct a zkPoA scheme that is derived from the inference of individual samples to attest to the effectiveness of the committed model
 by demonstrating its accuracy over public dataset  $\mathcal{D}=(\vect{x}_1,\dots,\vect{x}_M)$ 
with ground truth labels $\vect{T} = (t_1, \dots, t_M)$.
zkPoA requires the server to commit to a model with claimed accuracy on public sources. Once the model is committed and zkPoA is generated, it cannot be altered. The server has to use the model that has been committed previously for the successive inference tasks. 
Let $\vect{Y} = (y_1, \dots, y_M)$ be the predicted labels of $\mathcal{D}$, where $y_i \gets \mathsf{DPS}(\mlparam,\vect{x}_i)$ for $i \in [1,M]$.
The accuracy of MLIP model over $\mathcal{D}$ is 
%
$	\psi = \frac{\Sigma_{i=1}^M (y_i \stackrel{?}{=} t_i)}{M}$ 
where $0 \le \psi \le 1$. 

In our zkPoA, it suffices to show the committed model maintains \emph{at least} $ \psi $ accuracy (rather than the precise number) by proving that at least $\psi \cdot M $ samples are classified correctly.
This reduces the complexity since the prover does not have to prove some samples are misclassified (which incurs complex circuits for proof of inequality).
Our zkPoA is as follows.

We expand $\vect{Y}$ and $\vect{T}$ to value-index pairs as $\vect{Y} = \{(y_1, 1), \dots, (y_M, M)\}$, $\vect{T} = \{(t_1, 1), \dots, (t_M, M)\}$. 
The prover shuffles $\vect{Y}$ and $\vect{T}$ to 
$\vect{Y}'$ and $\vect{T}'$ using permutation functions $\sigma_1$, $\sigma_2$, respectively, which have two goals:
$ (i) $ hide which samples are classified correctly, 
and $ (ii) $ reduce the computation cost by rearranging correctly classified samples as first items in $\vect{Y}'$ and $\vect{T}'$.  
Therefore, $\Prover$ needs to prove: 
$ (i) $ first  $ \psi \cdot M $ items in $\vect{Y}'$ and $\vect{T}'$ are identical, 
$ (ii) $ $\vect{Y}'$ (resp. $\vect{T}'$ ) is a permutation of $\vect{Y}$ (resp. $\vect{T}$), 
and $(iii)$ two permutations are the same.

Suppose the permuted sets are $\vect{Y}' = \{(y'_1, \sigma_1(1)), \dots, (y'_M, \sigma_1(M))\}$ and $\vect{T}' = \{(t'_1, \sigma_2(1)),\dots, (t'_M, \sigma_2(M))\}$, where $y'_i = y_{\sigma_1(i)}$ and $t'_i = t_{\sigma_2(i)}$. 
The prover provides $\vect{Y}'$ and $\vect{T}'$ as the auxiliary witnesses.
Let $\xi$ be a random challenge chosen by the verifier.
To perform the permutation test, $\Prover$ computes intermediate values $\vect{\tilde{Y}} = \{\tilde{y}_i\}, \vect{\bar{Y}} = \{\bar{y}_i\}, \vect{\tilde{T}} = \{\tilde{t}_i\}$ and $\vect{\bar{T}} = \{\bar{t}_i\}$ such that for each $i \in [1, M]$:

\vspace{-.7em}
\begin{equation*}
	\begin{aligned}
		\tilde{y}_i & = y_i + \xi \cdot i & \text{and} \;\; \bar{y}_i & = y'_i + \xi \cdot \sigma_1(i) \\
		\tilde{t}_i & = t_i + \xi \cdot i & \text{and} \;\; \bar{t}_i & = t'_i + \xi \cdot \sigma_2(i)
	\end{aligned}
\end{equation*}
	
\noindent The set of constraints for our zkPoA includes all the constraints to prove each $y_i$ plus the following constraints

	\begin{equation*}\label{eq:cons:acc}
		\begin{cases}
			y'_i - t'_i = 0 \text{ for } i \in [1, \psi \cdot M]\\
			\sigma_1(i) - \sigma_2(i) = 0 \text{ for } i \in [1, M]\\
			\mathsf{Perm}(\vect{\tilde{Y}}, \vect{\bar{Y}}) \\
			\mathsf{Perm}(\vect{\tilde{T}}, \vect{\bar{T}}) 
		\end{cases}
	\end{equation*}

\begin{figure}[t]
\small
\begin{mdframed}
	\begin{Alg}[$y \gets \mathsf{DPS}(\mlparam,\vect{x})$] \label{alg:DPS}
		\textbf{Input}: Data sample $ \vect{x} \in \F^m $, MLIP model parameters $ \mlparam = (\vect{h}, \vect{g}, \vect{\bar{h}}, \vect{\bar{g}}, \eta, \vect{\bar{x}}, \vect{V}', \{\vect{x}_i, \delta_i^{(\hat{c} )},b^{(\hat{c} )}\}_{i\in \set{I}^{(\hat{c} )}, \hat{c} \in [1,s]}, \gamma) $\\
		\textbf{Output}: Inference result $ y $.
		\begin{algorithmic}[1]
			\For {$ \ell=1 $ to $ d $} \label{alg:DPS:DWT:start}
			\State $t_\ell \gets \frac{m}{2^\ell}$ and $t'_\ell \gets \frac{m}{2^{\ell-1}}$
			\For {$ i =1   $ to $ t_\ell $}
				\State $\vect{z}_{\ell}[i]  \gets  \sum_{j=1}^{c} \vect{h}[j] \cdot \vect{z}_{\ell-1}[(2i+j-2)_{\text{mod }t'_\ell}]$
				\State $\vect{z}_{\ell}[i+t_{\ell}] \gets \sum_{j=1}^{c} \vect{g}[j] \cdot \vect{z}_{\ell-1}[(2i+j-2)_{\text{mod }t'_{\ell}}]$
			\EndFor
			\For {$ i =1   $ to $ t_\ell $}
			\State $ \vect{z}'_{\ell}[i] \gets \vect{z}_{\ell}[i]$
			\If{$\left|\vect{z}_{\ell}[{i+t_\ell}]\right| - \eta > 0$}
			\State $\vect{z}'_{\ell}[{i+t_\ell}] \gets \mathsf{sign}(\vect{z}_{\ell}[{i+t_\ell}]) (\vect{z}_{\ell}[{i+t_\ell}] - \eta)$
			\Else 
			\State $\vect{z}'_{\ell}[{i+t_\ell}] \gets 0$
			\EndIf
			\EndFor
			\For {$ i =1   $ to $ t_\ell $}
			\State $\vect{\hat{x}}_\ell[2i-1] \gets \sum\limits_{j=1}^{c/2}(\vect{\bar{h}}[2j-1]\cdot \vect{z}'_\ell[(i+j-1)_{\text{mod }t_\ell}]$
			\Statex \quad \quad\quad$+\vect{\bar{h}}[2j]\cdot \vect{z}'_\ell[t_\ell+(i+j-1)_{\text{mod }t_\ell}])$
			\State $	\vect{\hat{x}}_\ell[2i] \gets \sum\limits_{j=1}^{c/2}(\vect{\bar{g}}[2j-1]\cdot \vect{z}'_\ell[(i+j-1)_{\text{mod }t_\ell}]$
			\Statex \quad \quad\quad $+\vect{\bar{g}}[2j]\cdot \vect{z}'_\ell[t_\ell+(i+j-1)_{\text{mod }t_\ell}])$
			\EndFor
			\EndFor		\label{alg:DPS:DWT:end}
			\State $ \vect{\tilde{x}} \gets (\vect{\hat{x}}_d - \vect{\bar{x}}) \mat{V}'  $ \label{alg:DPS:PCA}
			\For {$ \hat{c} = 1 $ to $ s $} \label{alg:DPS:SVM:start}
				\State Let $ \set{I}^{(\hat{c} )} =  \{ i: \delta^{(\hat{c} )}_i > 0 \} $
				\State $ y_{\hat{c}}  \gets \sum_{i\in \set{I}^{(\hat{c})}} \delta_i^{(\hat{c})} y^{(\hat{c})}_i \phi(\vect{\tilde{x}}, \vect{x}_i)+ b^{(\hat{c})}$
			\EndFor
			\State $ y_c \gets \max (y_1,\dots,y_s)$  \label{alg:DPS:SVM:end}
			\State		\Return $ c $
		\end{algorithmic}
	\end{Alg}
\end{mdframed}\vspace{-0.7em}
	\caption{MLIP with DWT, PCA and SVM algorithms.}\label{alg:MLIP:DWTPCASVM}\vspace{-1em}
\end{figure}
\vspace{-.11em}
\section{Analysis}\label{sec:analysis}
\PP{Complexity}
Let $ m, k$ be the dimensions of the raw data sample and the feature vector by PCA, respectively. Let $ s, t $ be the number of SVM classes and the number of support vectors for all classes, respectively. In DWT, our scheme requires $ 8\log_2\frac{2m}{c} $  constraints for DWT decomposition \autoref{eq:dwt:decompose:final} and reconstruction \autoref{eq:dwt:reconstruct:prove:final}, while the thresholding \autoref{eq:dwt:threshold} incurs $(3n+9)(m-\frac{c}{2})$ constraints, where $ n $ is the size (in bits) of each value per dimension of the raw data sample, $c$ is the dimension of the high-pass and low-pass filters. In total, our scheme requires $16\log_2 \frac{2m}{c} +(3n+9)(m-\frac{c}{2})$ constraints for proving DWT. In PCA, the number of constraints is $ m $ \autoref{eq:pca:finalconstraints}. This is reduced from $ mk $ compared with direct proving \autoref{eq:PCA} due to random linear combination. In SVM classification \autoref{eq:svm:final}, our scheme incurs $ {(2n+k)t+2s} $ constraints for proving RBF kernel projection, and $ {(3n+6)(s-1)+2s} $ constraints for proving the classification for $ s $ classes and $ t $ constraints for the final decision function. The permutation trick in our proposed $\mathsf{Max}$ gadget permits us to reduce the number of comparisons from $O(s^2)$ in generic circuits to $O(s)$ . In total, our scheme incurs $ (2n+k)t + 4s + (3n+6)(s-1) $ constraints for proving $ s $-class SVM classification with RBF kernel.
\autoref{tab:complexity} summarizes the complexity of our framework, compared with directly proving DWT, PCA, and SVM computations with generic circuits.

For zkPoA, suppose the number of samples in the testing dataset is $M$, and proving one testing data incurs $N$ constraints. Therefore, our zkPoA incurs $(N+4)M$ constraints for proving the accuracy.  

\begin{table}[t!]
	\caption{Complexity of \sys~vs. generic circuit (baseline).}
	\label{tab:complexity}
	\centering
	\resizebox{.7\columnwidth}{!}{
	\begin{tabular}{|cc| c | c|}
		\hline
		\multicolumn{2}{|l|}{}                     & \textbf{\sys}                                       & \textbf{Generic circuit}                             \\ \hline
		\multicolumn{1}{|c|}{\multirow{3}{*}{DWT}} & Decomposition         & $8\log_2\frac{2m}{c}$                                          & $8m-4c$                                \\
		\multicolumn{1}{|c|}{}                     & Thresholding      & $(3n+9)(m-\frac{c}{2})$                         & $(5n+12)(m-\frac{c}{2})$                 \\
		\multicolumn{1}{|c|}{}                     & Reconstruction        & $8\log_2\frac{2m}{c}$                                          & $8m-4c$                                \\ \hline
		\multicolumn{2}{|c|}{PCA}                                          & $m$                                        & $mk$                                \\ \hline
		\multicolumn{2}{|c|}{Multi-class SVM} & $(2n+k)t + 2s $ & $(2n+k+2)t +s$ \\ 
		
		\multicolumn{2}{|c|}{(w/ RBF)} & $ +(3n+6)(s-1)+2s $ & $+(s^2-s)(2n+5)+2s-2$ \\ \hline
	\end{tabular}
}
\end{table}

\PP{Security}
We analyze the security of our scheme. Specifically, we have the following theorem.
\begin{Theorem} \label{theorem:sys:security}
	Our \sys~scheme in  \autoref{pro:sys} is a zero-knowledge MLIP as defined in \autoref{def:sec:zkmlip} given that the backend CP-ZKP is secure by \autoref{theorem:spartan}.
\end{Theorem}

\begin{proof}
	See \autoref{sec:proveMainTheorem}
\end{proof}

\section{Implementation}\label{sec:impl}
We fully implemented our proposed framework in Python and Rust, consisting of approximately 2,500 lines of code in total.
For DWT, we implemented the Daubechies DB4 algorithm \cite{vonesch2007generalized}.
We used $\mathtt{sklearn}$ \cite{scikit-learn} to implement the training phase of PCA and SVM.
On the other hand, we implemented the inference phase of PCA and SVM from scratch to obtain all the witnesses for generating the proofs.
We used fixed-point number representation for all the values being processed in our framework.
Each value can be represented by 64 bits, which reserves 1 bit for the sign, 31 bits for the integer part, and 32 bits for the fractional part. 
 We used the exponent gadget to prove the RBF kernel of the form $e^{\gamma || \vect{x}_i - \vect{x}_j||^2}$, where the base $e^{\gamma}$ is public and the exponent $|| \vect{x}_i - \vect{x}_j||^2$ is secret (witness).
 As shown in \autoref{sec:function}, our gadget precomputes $a^{2^{i-1}}$, where $a = e^{-\gamma}$ and $i$ is the index of the binary representation of the exponent. 
 We used a fixed-point arithmetic to represent the exponent.
 Since it suffices to set $\gamma=10^{-3}$ for RBF kernel, we used 20 bits to represent the fractional part of the exponent, which suffices to cover most of the cases in our test set.
 There are few samples that cause the fractional part of the exponent to exceed ${20}$ bits.
 In this case, we truncated the fractional part of the witness that exceeds $20$ bits, leading to a small accuracy loss (see \autoref{sec:exp:accuracyloss}).

In our implementation, we transformed the arithmetic constraints and the witnesses generated from ML algorithms into R1CS relations using the compact encoding method in $\mathtt{libspartan}$ \cite{spartanAPI} and then invoked its library APIs to create proofs and verification.
Concretely, we used $\mathtt{Splartan}_{\mathtt{DL}}$ scheme, which implements \textit{(i)} Hyrax polynomial commitment \cite{wahby2018doubly}, \textit{(ii)}
$\texttt{curve25519-dalek}$ \cite{curve25519} for curve arithmetic in prime order ristretto group, \textit{(iii)} a separate dot-product proof protocol for each round of the sum-check protocol for zero-knowledge property, and  \textit{(iv)} \texttt{merlin} \cite{merlin} for non-interactive proof via Fiat-Shamir transformation.

Our implementation is available at \href{https://github.com/vt-asaplab/ezDPS}{https://github.com/vt-asaplab/ezDPS}.

\section{Experimental Evaluation}\label{sec:exp}
\subsection{Configuration}
\noindent \textbf{Hardware}. We ran all the experiments on a 2020 Macbook Pro, which was equipped with a 2.0 GHz 4-core Intel Core i5 CPU, 16GB DDR4 RAM. 
Currently, we did not make use of thread-level parallelization to accelerate the proving/verification time.
The experimental results reported in this section are with single-thread computation, which can be further improved once multi-thread parallelization is employed.

\PP{Dataset} We evaluated our scheme on three public datasets, 
including the ECG dataset in UCR Time Series Classification Archive (UCR-ECG)\cite{UCRArchive2018},
 Labeled Faces in the Wilds (LFW) \cite{huang2008labeled}, and Cifar-100 \cite{krizhevsky2009learning}. 
UCR-ECG contains 1800 records of ECG signals, each being of length 750.
LFW contains 5749 human faces, where each image is of size $125 \times 94$ bits.
Cifar-100 contains 100 classes, and the dimension of the samples is 3072. 
We used the subset of each dataset for the different number of classes. 

\PP{Parameters} 
We used standard parameters as suggested in {Spartan} \cite{setty2020spartan} (e.g., curve25519) for 128-bit security.
We evaluated the performance of our proposed methods with varied numbers of classes ($ s $) and PCA dimensions ($ k $) (see \autoref{tab:model_info}).
For LFW dataset, we scaled the dimension of the image inputs to 4200 when the number of classes is small (i.e., 8 and 16), and to 5655 for many classes ($ > 32 $).
For DWT processing, 
we set the number of recursion levels to be $ 1 $ for noise reduction 
and $ \eta=0.2 $ for processing the detail coefficients. 
For PCA, we selected the number of eigenvectors $ k $ such that they can capture at least 90\% of the variance.
We presented the concrete number of $ k $ w.r.t different sizes of the datasets in \autoref{tab:model_info}. 
Finally, we used the Grid Search method to find the best parameters for SVM and set $C=1$, $\gamma=0.001$.


\begin{table}[t]
	\centering
    \small
	\caption{{Detailed model parameters.}}
	\label{tab:model_info}
		\begin{threeparttable}
			\begin{tabular}{|cccc|cccc|clcl|}
				\hline
				\multicolumn{4}{|c|}{\textbf{UCR-ECG}}  & \multicolumn{4}{c|}{ \textbf{Cifar-100}} & \multicolumn{4}{c|}{\textbf{LFW}} \\ \cline{1-4} \cline{5-8} \cline{9-12} 
				$ m $ & $ k $ &  $ s $ & $ t $ &  $ m $ & $ k $ & $ s $ & $ t $ & \textbf{$ m $} & $ k $ & $ s $ & $ t $ \\ \hline
				750 & 33 & 4 & 54 &  3072 & 98 & 4 & 676 &5655 & 119 & 8 & 1005 \\
				750 & 34 & 8 & 115 & 3072 & 108 & 8 & 1967 &5655 & 121 & 16 & 1236 \\
				750 & 57 & 16 & 317 &  3072 & 121 & 16 & 2950 &5655 & 123 & 32 & 1533 \\
				750 & 55 & 32 & 795 &  3072 & 120 & 32 & 3354 &5655 & 125 & 64 & 1718 \\
				750 & 47 & 42 & 1061 &  3072 & 112 & 64 & 4627 &5655 & 120 & 128 & 1384 \\
				&  &  &  & 3072 & 108 & 100 & 6623 &5655 & 106 & 256 & 4895\\
				&  &  &  & & & & &5655 & 102 & 512 & 3862 \\
				&  &  &  & & & & &5655 & 121 & 1024 & 3233 \\
				& & & & & & & &5655 & 118 & 2048 & 2645   \\ \hline
			\end{tabular}
			\begin{tablenotes}
				\begin{footnotesize}
					\item $ m $: dimension of raw data, $ k $: dimension of feature vector by PCA, $ s $: number of distinct class labels, $ t $: number of all support vectors in all classes.	
				\end{footnotesize}	
			\end{tablenotes}
		\end{threeparttable}
\end{table}

\begin{figure*}[!t]
	\centering
	\captionsetup[subfigure]{justification=centering}
	\begin{subfigure}{1\textwidth}	
		\resizebox{1\textwidth}{!}{
			\begin{subfigure}{.3\textwidth}
%
%
\definecolor{A}{HTML}{e6194B}%
\definecolor{B}{HTML}{f58231}%
\definecolor{C}{HTML}{4363d8}%
\definecolor{D}{HTML}{911eb4}%
\definecolor{E}{HTML}{3cb44b}%
\definecolor{F}{rgb}{0.92900,0.69400,0.12500}%
\definecolor{G}{HTML}{808000}%
\definecolor{H}{HTML}{000000}%
\begin{tikzpicture}
	
	\begin{axis}[%
		width=.8\textwidth,
		height=.5\textwidth,
		at={(1.128in,0.894in)},
		scale only axis,
		xmin=0,
		xmax=4.05,
		xlabel={\# classes ($ s $)},
		xtick={0,1,2,3,4},
		xticklabels={$2^2$, $ 2^3 $, $ 2^4 $, $ 2^5 $, $ 2^6 $},
		ymin=-200,
		ymax=3200,
		ytick distance=1000,
		ylabel = {\Prover~time (sec)},
		ylabel shift=-5pt,
		yticklabel shift={0cm},
		axis background/.style={fill=white},
		legend columns=3,
		legend style={legend cell align=left, align=left, fill=none, draw=none,inner sep=-0pt, row sep=0pt},
		legend pos = north west,
		ymajorgrids,
		xmajorgrids,
		grid style={line width=.5pt, draw=gray!90,dashed},
		major grid style={line width=.2pt,draw=gray!50},
		minor y tick num=5,
		]
		\addplot [color=A, solid, mark=diamond*, mark options={solid, A}]
		table[row sep=crcr]{%
			0 321\\
			1 372\\
			2 451\\
			3 505\\
			3.31 512\\
		};
		\addlegendentry{\sys}
		
		\addplot [color=B, dashed, mark=square*, mark options={solid, B}]
		table[row sep=crcr]{%
			0 1429\\
			1 1665\\
			2 2223\\
			3 2631\\
			3.31 2799\\
		};
		\addlegendentry{Baseline}		
	\end{axis}
\end{tikzpicture}%
			\end{subfigure}\hspace{4mm}
			\begin{subfigure}{.3\textwidth}
%
%
\definecolor{A}{HTML}{e6194B}%
\definecolor{B}{HTML}{f58231}%
\definecolor{C}{HTML}{4363d8}%
\definecolor{D}{HTML}{911eb4}%
\definecolor{E}{HTML}{3cb44b}%
\definecolor{F}{rgb}{0.92900,0.69400,0.12500}%
\definecolor{G}{HTML}{808000}%
\definecolor{H}{HTML}{000000}%
\begin{tikzpicture}
	
	\begin{axis}[%
		width=.8\textwidth,
		height=.5\textwidth,
		at={(1.128in,0.894in)},
		scale only axis,
		xmin=0,
		xmax=4.05,
		xlabel={\# classes ($ s $)},
		xtick={0,1,2,3,4},
		xticklabels={$2^2$, $ 2^3 $, $ 2^4 $, $ 2^5 $, $ 2^6 $},
		ymin=-2,
		ymax=40,
		ytick distance=10,
		ylabel = {\Verifier~time (sec)},
		ylabel shift=-5pt,
		yticklabel shift={0cm},
		axis background/.style={fill=white},
		legend columns=3,
		legend style={legend cell align=left, align=left, fill=none, draw=none,inner sep=-0pt, row sep=0pt},
		legend pos = north west,
		ymajorgrids,
		xmajorgrids,
		grid style={line width=.5pt, draw=gray!90,dashed},
		major grid style={line width=.2pt,draw=gray!50},
		minor y tick num=5,
		]
		\addplot [color=A, solid, mark=diamond*, mark options={solid, A}]
		table[row sep=crcr]{%
			0 2.39\\
			1 2.91\\
			2 3.87\\
			3 4.479\\
			3.31 4.568\\
		};
		\addlegendentry{\sys}
		
		\addplot [color=B, dashed, mark=square*, mark options={solid, B}]
		table[row sep=crcr]{%
			0 13.833\\
			1 16.532\\
			2 22.869\\
			3 27.623\\
			3.31 29.565\\
		};
		\addlegendentry{Baseline}

	\end{axis}
\end{tikzpicture}%
			\end{subfigure}
			\begin{subfigure}{.3\textwidth}
%
%
\definecolor{A}{HTML}{e6194B}%
\definecolor{B}{HTML}{f58231}%
\definecolor{C}{HTML}{4363d8}%
\definecolor{D}{HTML}{911eb4}%
\definecolor{E}{HTML}{3cb44b}%
\definecolor{F}{rgb}{0.92900,0.69400,0.12500}%
\definecolor{G}{HTML}{808000}%
\definecolor{H}{HTML}{000000}%
\begin{tikzpicture}
	
	\begin{axis}[%
		width=.8\textwidth,
		height=.5\textwidth,
		at={(1.128in,0.894in)},
		scale only axis,
		xmin=0,
		xmax=4.05,
		xlabel={\# classes ($ s $)},
		xtick={0,1,2,3,4},
		xticklabels={$2^2$, $ 2^3 $, $ 2^4 $, $ 2^5 $, $ 2^6 $},
		ymin=-0000,
		ymax=20500,
		ytick distance=5000,
		ylabel = {Proof size (KB)},
		ylabel shift=-5pt,
		yticklabel shift={0cm},
		axis background/.style={fill=white},
		legend columns=3,
		legend style={legend cell align=left, align=left, fill=none, draw=none,inner sep=-0pt, row sep=0pt},
		legend pos = north west,
		ymajorgrids,
		xmajorgrids,
		grid style={line width=.5pt, draw=gray!90,dashed},
		major grid style={line width=.2pt,draw=gray!50},
		minor y tick num=5,
		]
		\addplot [color=A, solid, mark=diamond*, mark options={solid, A}]
		table[row sep=crcr]{%
			0 1085\\
			1 1309\\
			2 1717\\
			3 1973\\
			3.31 2010\\
		};
		\addlegendentry{\sys}
		
		\addplot [color=B, dashed, mark=square*, mark options={solid, B}]
		table[row sep=crcr]{%
			0 9210\\
			1 10921\\
			2 14909\\
			3 17117\\
			3.31 17523\\
		};
		\addlegendentry{Baseline}

	\end{axis}
\end{tikzpicture}%
			\end{subfigure}
		}\vspace{-.7em}
		\caption{UCR-ECG}\label{fig:exp:IM} 
	\end{subfigure}\\
	\begin{subfigure}{1\textwidth}
		\resizebox{1\textwidth}{!}{
			\begin{subfigure}{.3\textwidth}
%
%
\definecolor{A}{HTML}{e6194B}%
\definecolor{B}{HTML}{f58231}%
\definecolor{C}{HTML}{4363d8}%
\definecolor{D}{HTML}{911eb4}%
\definecolor{E}{HTML}{3cb44b}%
\definecolor{F}{rgb}{0.92900,0.69400,0.12500}%
\definecolor{G}{HTML}{808000}%
\definecolor{H}{HTML}{000000}%
\begin{tikzpicture}
	\begin{axis}[%
		width=.8\textwidth,
		height=.5\textwidth,
		at={(1.128in,0.894in)},
		scale only axis,
		xmin=0,
		xmax=5.5,
		xlabel={\# classes ($ s $)},
		xtick={0,1,2,3,4,5},
		xticklabels={$2^2$, $ 2^3 $, $ 2^4 $, $ 2^5 $, $ 2^6 $, $ 2^7 $},
		ymin=-500,
		ymax=15000,
		ytick distance=4000,
		ylabel = {\Prover~time (sec)},
		ylabel shift=-5pt,
		yticklabel shift={0cm},
		axis background/.style={fill=white},
		legend columns=3,
		legend style={legend cell align=left, align=left, fill=none, draw=none,inner sep=-0pt, row sep=0pt},
		legend pos = north west,
		ymajorgrids,
		xmajorgrids,
		grid style={line width=.5pt, draw=gray!90,dashed},
		major grid style={line width=.2pt,draw=gray!50},
		minor y tick num=5,
		]
		\addplot [color=A, solid, mark=diamond*, mark options={solid, A}]
		table[row sep=crcr]{%
			0 766\\
			1 816\\
			2 915\\
			3 934\\
			4 1173\\
			4.64 1449\\
		};
		\addlegendentry{\sys}
		
		\addplot [color=B, dashed, mark=square*, mark options={solid, B}]
		table[row sep=crcr]{%
			0 3282\\
			1 4292\\
			2 5799\\
			3 6748\\
			4 10175\\
			4.64 14328\\
		};
		\addlegendentry{Baseline}

	\end{axis}
\end{tikzpicture}%
			\end{subfigure}\hspace{4mm}
			\begin{subfigure}{.3\textwidth}
%
%
\definecolor{A}{HTML}{e6194B}%
\definecolor{B}{HTML}{f58231}%
\definecolor{C}{HTML}{4363d8}%
\definecolor{D}{HTML}{911eb4}%
\definecolor{E}{HTML}{3cb44b}%
\definecolor{F}{rgb}{0.92900,0.69400,0.12500}%
\definecolor{G}{HTML}{808000}%
\definecolor{H}{HTML}{000000}%
	\begin{tikzpicture}
		\begin{axis}[%
			width=.8\textwidth,
			height=.5\textwidth,
			at={(1.128in,0.894in)},
			scale only axis,
			xmin=0,
			xmax=5.5,
			xlabel={\# classes ($ s $)},
			xtick={0,1,2,3,4,5},
			xticklabels={$2^2$, $ 2^3 $, $ 2^4 $, $ 2^5 $, $ 2^6 $, $ 2^7 $},
			ymin=-.500,
			ymax=17.000,
			ytick distance=4.000,
			ylabel = {\Verifier~time (sec)},
			ylabel shift=-5pt,
			yticklabel shift={0cm},
			axis background/.style={fill=white},
			legend columns=3,
			legend style={legend cell align=left, align=left, fill=none, draw=none,inner sep=-0pt, row sep=0pt},
			legend pos = north west,
			ymajorgrids,
			xmajorgrids,
			grid style={line width=.5pt, draw=gray!90,dashed},
			major grid style={line width=.2pt,draw=gray!50},
			minor y tick num=5,
			]
		\addplot [color=A, solid, mark=diamond*, mark options={solid, A}]
		table[row sep=crcr]{%
			0 3.177\\
			1 3.402\\
			2 3.735\\
			3 3.779\\
			4 4.254\\
			4.64 4.672\\
		};
		\addlegendentry{\sys}
		
		\addplot [color=B, dashed, mark=square*, mark options={solid, B}]
		table[row sep=crcr]{%
			0 7.389\\
			1 8.657\\
			2 10.116\\
			3 11.165\\
			4 13.847\\
			4.64 16.110\\
		};
		\addlegendentry{Baseline}

	\end{axis}
\end{tikzpicture}%
			\end{subfigure}
			\begin{subfigure}{.3\textwidth}
%
%
\definecolor{A}{HTML}{e6194B}%
\definecolor{B}{HTML}{f58231}%
\definecolor{C}{HTML}{4363d8}%
\definecolor{D}{HTML}{911eb4}%
\definecolor{E}{HTML}{3cb44b}%
\definecolor{F}{rgb}{0.92900,0.69400,0.12500}%
\definecolor{G}{HTML}{808000}%
\definecolor{H}{HTML}{000000}%
	\begin{tikzpicture}
		\begin{axis}[%
			width=.8\textwidth,
			height=.5\textwidth,
			at={(1.128in,0.894in)},
			scale only axis,
			xmin=0,
			xmax=5.5,
			xlabel={\# classes ($ s $)},
			xtick={0,1,2,3,4,5},
			xticklabels={$2^2$, $ 2^3 $, $ 2^4 $, $ 2^5 $, $ 2^6 $, $ 2^7 $},
			ymin=-000,
			ymax=10000,
			ytick distance=2000,
			ylabel = {Proof size (KB)},
			ylabel shift=-5pt,
			yticklabel shift={0cm},
			axis background/.style={fill=white},
			legend columns=3,
			legend style={legend cell align=left, align=left, fill=none, draw=none,inner sep=-0pt, row sep=0pt},
			legend pos = north west,
			ymajorgrids,
			xmajorgrids,
			grid style={line width=.5pt, draw=gray!90,dashed},
			major grid style={line width=.2pt,draw=gray!50},
			minor y tick num=5,
			]
		\addplot [color=A, solid, mark=diamond*, mark options={solid, A}]
		table[row sep=crcr]{%
			0 1804\\
			1 1895\\
			2 2028\\
			3 2041\\
			4 2243\\
			4.64 2410\\
		};
		\addlegendentry{\sys}
		
		\addplot [color=B, dashed, mark=square*, mark options={solid, B}]
		table[row sep=crcr]{%
			0 4687\\
			1 5441\\
			2 6383\\
			3 6924\\
			4 8315\\
			4.64 9338\\
		};
		\addlegendentry{Baseline}

	\end{axis}
\end{tikzpicture}%
			\end{subfigure}
		}\vspace{-.7em}
		\caption{{Reduced Cifar-100}}
		\label{fig:exp:cifar}
	\end{subfigure}\\
	\begin{subfigure}{1\textwidth}
		\resizebox{1\textwidth}{!}{
			\begin{subfigure}{.3\textwidth}
%
%
\definecolor{A}{HTML}{e6194B}%
\definecolor{B}{HTML}{f58231}%
\definecolor{C}{HTML}{4363d8}%
\definecolor{D}{HTML}{911eb4}%
\definecolor{E}{HTML}{3cb44b}%
\definecolor{F}{rgb}{0.92900,0.69400,0.12500}%
\definecolor{G}{HTML}{808000}%
\definecolor{H}{HTML}{000000}%
\begin{tikzpicture}

\begin{axis}[%
width=.8\textwidth,
height=.5\textwidth,
at={(1.128in,0.894in)},
scale only axis,
xmin=0,
xmax=8.5,
xlabel={\# classes ($ s $)},
xtick={0,1,2,3,4,5,6, 7, 8},
xticklabels={$ 2^3 $, $ 2^4 $, $ 2^5 $, $ 2^6 $, $ 2^7 $, $ 2^{8} $, $ 2^{9} $, $ 2^{10} $, $ 2^{11} $},
ymin=-200000,
ymax=2500000,
ytick distance=500000,
ylabel = {\Prover~time (sec)},
ylabel shift=-5pt,
yticklabel shift={0cm},
axis background/.style={fill=white},
legend columns=3,
legend style={legend cell align=left, align=left, fill=none, draw=none,inner sep=-0pt, row sep=0pt},
legend pos = north west,
ymajorgrids,
xmajorgrids,
grid style={line width=.5pt, draw=gray!90,dashed},
major grid style={line width=.2pt,draw=gray!50},
minor y tick num=5,
]
\addplot [color=A, solid, mark=diamond*, mark options={solid, A}]
  table[row sep=crcr]{%
  	0 1702\\
  	1 2549\\
  	2 3048\\
  	3 3405\\
  	4 3407\\
  	5 4893\\
  	6 5251\\
  	7 5667\\
  	8 6971\\
};
\addlegendentry{\sys}

\addplot [color=B, dashed, mark=square*, mark options={solid, B}]
table[row sep=crcr]{%
0 11491\\
1 21237\\
2 21676\\
3 20321\\
4 27095\\
5 57426\\
6 170484\\
7 623545\\
8 2439807\\
};
\addlegendentry{Baseline}

\end{axis}
\end{tikzpicture}%
			\end{subfigure}\hspace{4mm}
			\begin{subfigure}{.3\textwidth}
%
%
\definecolor{A}{HTML}{e6194B}%
\definecolor{B}{HTML}{f58231}%
\definecolor{C}{HTML}{4363d8}%
\definecolor{D}{HTML}{911eb4}%
\definecolor{E}{HTML}{3cb44b}%
\definecolor{F}{rgb}{0.92900,0.69400,0.12500}%
\definecolor{G}{HTML}{808000}%
\definecolor{H}{HTML}{000000}%
\begin{tikzpicture}
	
\begin{axis}[%
	width=.8\textwidth,
	height=.5\textwidth,
	at={(1.128in,0.894in)},
	scale only axis,
	xmin=0,
	xmax=8.5,
	xlabel={\# classes ($ s $)},
	xtick={0,1,2,3,4,5,6, 7, 8},
	xticklabels={$ 2^3 $, $ 2^4 $, $ 2^5 $, $ 2^6 $, $ 2^7 $, $ 2^{8} $, $ 2^{9} $, $ 2^{10} $, $ 2^{11} $},
	ymin=-10.000,
	ymax=130,
	ytick distance=20.000,
	ylabel = {\Verifier~time (sec)},
	ylabel shift=-5pt,
	yticklabel shift={0cm},
	axis background/.style={fill=white},
	legend columns=3,
	legend style={legend cell align=left, align=left, fill=none, draw=none,inner sep=-0pt, row sep=0pt},
	legend pos = north west,
	ymajorgrids,
	xmajorgrids,
	grid style={line width=.5pt, draw=gray!90,dashed},
	major grid style={line width=.2pt,draw=gray!50},
	minor y tick num=5,
	]
		\addplot [color=A, solid, mark=diamond*, mark options={solid, A}]
		table[row sep=crcr]{%
0 5.094\\
1 6.006\\
2 6.657\\
3 6.951\\
4 6.950\\
5 7.956\\
6 7.557\\
7 8.397\\
8 9.055\\
		};
		\addlegendentry{\sys}
		
		\addplot [color=B, dashed, mark=square*, mark options={solid, B}]
		table[row sep=crcr]{%
			0 14.065\\
			1 18.709\\
			2 19.409\\
			3 19.401\\
			4 22.182\\
			5 29.235\\
			6 42.497\\
			7 69.823\\
			8 123.600\\
		};
		\addlegendentry{Baseline}

	\end{axis}
\end{tikzpicture}%
			\end{subfigure}\hspace{4mm}
			\begin{subfigure}{.3\textwidth}
%
%
\definecolor{A}{HTML}{e6194B}%
\definecolor{B}{HTML}{f58231}%
\definecolor{C}{HTML}{4363d8}%
\definecolor{D}{HTML}{911eb4}%
\definecolor{E}{HTML}{3cb44b}%
\definecolor{F}{rgb}{0.92900,0.69400,0.12500}%
\definecolor{G}{HTML}{808000}%
\definecolor{H}{HTML}{000000}%
\begin{tikzpicture}

\begin{axis}[%
	width=.8\textwidth,
	height=.5\textwidth,
	at={(1.128in,0.894in)},
	scale only axis,
	xmin=0,
	xmax=8.5,
	xlabel={\# classes ($ s $)},
	xtick={0,1,2,3,4,5,6, 7, 8},
	xticklabels={$ 2^3 $, $ 2^4 $, $ 2^5 $, $ 2^6 $, $ 2^7 $, $ 2^{8} $, $ 2^{9} $, $ 2^{10} $, $ 2^{11} $},
	ymin=-000,
	ymax=60000,
	ytick distance=10000,
	ylabel = {Proof size (KB)},
	ylabel shift=-5pt,
	yticklabel shift={0cm},
	axis background/.style={fill=white},
	legend columns=3,
	legend style={legend cell align=left, align=left, fill=none, draw=none,inner sep=-0pt, row sep=0pt},
	legend pos = north west,
	ymajorgrids,
	xmajorgrids,
	grid style={line width=.5pt, draw=gray!90,dashed},
	major grid style={line width=.2pt,draw=gray!50},
	minor y tick num=5,
	]
		\addplot [color=A, solid, mark=diamond*, mark options={solid, A}]
		table[row sep=crcr]{%
			0 2678\\
			1 3057\\
			2 3419\\
			3 3537\\
			4 3538\\
			5 3955\\
			6 3789\\
			7 4138\\
			8 4411\\
		};
		\addlegendentry{\sys}
		
		\addplot [color=B, dashed, mark=square*, mark options={solid, B}]
		table[row sep=crcr]{%
			0 8932\\
			1 11943\\
			2 12219\\
			3 11818\\
			4 12937\\
			5 15918\\
			6 21336\\
			7 33545\\
			8 56856\\
		};
		\addlegendentry{Baseline}

	\end{axis}
\end{tikzpicture}%
			\end{subfigure}
		}\vspace{-.5em}
		\caption{{Reduced LFW}}
		\label{fig:exp:LFW} 
	\end{subfigure} \vspace{-.7em}
	\caption{Performance of our scheme compared with the baseline.}\vspace{-.7em}
	\label{fig:exp:overall}
\end{figure*}

\noindent \textbf{Counterpart comparison}. 
To our knowledge, we are the first to propose a zero-knowledge MLIP.
There is also no prior work that suggests zero-knowledge proof 
for each of the ML algorithms (i.e., DWT, PCA, and SVM) in our framework.
Thus, we chose to compare with the na\"ive approach, in which we hardcore the whole DWT, PCA, and SVM computation into the circuit and ran the same CP-ZKP backend (i.e., Spartan).
%
%
We compared \sys~with this baseline to demonstrate our advantage in reducing the proving time, verification time, and proof size.
We also report the accuracy of $\sys$ to demonstrate the advantage of ML pipeline processing.

{%
\PP{Evaluation metrics} 
We assess the performance of our scheme and the baseline approach in terms of 
proving time, verification time and proof size (\autoref{sec:exp:overall} and \autoref{sec:exp:detail_cost}).
Note that for such cryptographic performance evaluation, we only used a reduced dataset of Cifar-100 and LFW that yield concrete model parameters after training as presented in \autoref{tab:model_info}.
For UCG-ECG, we used the whole set as it is already small.
We did not not evaluate on the whole set of Cifar-100 and LFW due to our limited hardware and the expensive cryptographic overhead incurred by the baseline.
Instead, we report the accuracy of plain ML techniques and estimate the performance of our scheme when tested on the whole dataset (\autoref{sec:exp:accuracyloss}).
}

\begin{figure*}[!t]
	\centering
	\captionsetup[subfigure]{justification=centering}
	\begin{subfigure}{1\textwidth}	
		\centering
		\resizebox{.99\textwidth}{!}{
			\begin{subfigure}{.31\textwidth}
				\definecolor{A}{HTML}{e6194B}%
\definecolor{B}{HTML}{f58231}%
\definecolor{C}{HTML}{4363d8}%
\definecolor{D}{HTML}{911eb4}%
\definecolor{E}{HTML}{3cb44b}%
\definecolor{F}{rgb}{0.92900,0.69400,0.12500}%
\definecolor{G}{HTML}{808000}%
\definecolor{H}{HTML}{000000}%

\begin{tikzpicture} 
	\begin{axis}[
		ybar stacked,
		xtick={0,1,2,3,4},
		xticklabels = {$2^2$, $2^3$, $2^4$, $2^5$, $42$},
		width=.8\textwidth,
		height=.4\textwidth,
		at={(1.128in,0.894in)},
		scale only axis,
		xlabel={\# classes ($ s $)},
		ylabel = {\Prover~time (sec)},
		ylabel shift=-7pt,
		yticklabel shift={0cm},
		axis background/.style={fill=white},
		legend columns=3,
		legend style={legend cell align=left, align=left, fill=none, draw=none,inner sep=-0pt, row sep=0pt,font=\tiny},
		legend pos = north west,
		ymajorgrids,
		xmajorgrids,
		ymin=0,
		grid style={line width=.5pt, draw=gray!90,dashed},
		major grid style={line width=.2pt,draw=gray!50},
		bar width=7pt,
		ymin=0,
		ymax=1000,
		legend columns=3,
		scaled y ticks=base 10:-3,
		]
		\addplot [blue,fill=white,pattern=north east lines,pattern color=.,]coordinates {
			(0, 160) (1, 161) (2, 165) (3, 164)(4, 166)
		};
		\addplot coordinates {
			(0, 18) (1, 18) (2, 18) (3, 17)(4, 19)
		};
		\addplot [black,fill=white,pattern=dots,pattern color=.,]coordinates {
			(0, 142) (1, 189) (2, 273) (3, 324)(4, 337)
		};
		\legend{DWT, PCA, SVM}
	\end{axis}
\end{tikzpicture}\vspace{-.5em}
			\end{subfigure}\hspace{4mm}
			\begin{subfigure}{.31\textwidth}
				\definecolor{A}{HTML}{e6194B}%
\definecolor{B}{HTML}{f58231}%
\definecolor{C}{HTML}{4363d8}%
\definecolor{D}{HTML}{911eb4}%
\definecolor{E}{HTML}{3cb44b}%
\definecolor{F}{rgb}{0.92900,0.69400,0.12500}%
\definecolor{G}{HTML}{808000}%
\definecolor{H}{HTML}{000000}%

\begin{tikzpicture} 
	\begin{axis}[
		ybar stacked,
		xtick={0,1,2,3,4},
		xticklabels = {$2^2$, $2^3$, $2^4$, $2^5$, $42$},
		width=.8\textwidth,
		height=.4\textwidth,
		at={(1.128in,0.894in)},
		scale only axis,
		xlabel={\# classes ($ s $)},
		ylabel = {\Verifier~time (sec)},
		ylabel shift=-7pt,
		yticklabel shift={0cm},
		axis background/.style={fill=white},
		legend columns=3,
		legend style={legend cell align=left, align=left, fill=none, draw=none,inner sep=-0pt, row sep=0pt,font=\tiny},
		legend pos = north west,
		ymajorgrids,
		xmajorgrids,
		ymin=0,
		ymax = 5.5,
		grid style={line width=.5pt, draw=gray!90,dashed},
		major grid style={line width=.2pt,draw=gray!50},
		bar width=7pt,
		legend columns=3,
		]
		\addplot  [blue,fill=white,pattern=north east lines,pattern color=.,]coordinates {
			(0, 0.471) (1, 0.472) (2, 0.482) (3, 0.483)(4, 0.485)
		};
		\addplot coordinates {
			(0, 0.198) (1, 0.198) (2, 0.211) (3, 0.213)(4, 0.211)
		};
		\addplot  [black,fill=white,pattern=dots,pattern color=.,] coordinates {
			(0, 1.659) (1, 2.2) (2, 3.183) (3, 3.803)(4, 3.895)
		};
		\legend{DWT, PCA, SVM}
	\end{axis}
\end{tikzpicture}\vspace{-.5em}
			\end{subfigure}
			\begin{subfigure}{.31\textwidth}
				\definecolor{A}{HTML}{e6194B}%
\definecolor{B}{HTML}{f58231}%
\definecolor{C}{HTML}{4363d8}%
\definecolor{D}{HTML}{911eb4}%
\definecolor{E}{HTML}{3cb44b}%
\definecolor{F}{rgb}{0.92900,0.69400,0.12500}%
\definecolor{G}{HTML}{808000}%
\definecolor{H}{HTML}{000000}%

\begin{tikzpicture} 
	\begin{axis}[
		ybar stacked,
		xtick={0,1,2,3,4},
		xticklabels = {$2^2$, $2^3$, $2^4$, $2^5$, $42$},
		width=.8\textwidth,
		height=.4\textwidth,
		at={(1.128in,0.894in)},
		scale only axis,
		xlabel={\# classes ($ s $)},
		ylabel = {Proof size (KB)},
		ylabel shift=-7pt,
		yticklabel shift={0cm},
		axis background/.style={fill=white},
		legend columns=3,
		legend style={legend cell align=left, align=left, fill=none, draw=none,inner sep=-0pt, row sep=0pt,font=\tiny},
		legend pos = north west,
		ymajorgrids,
		xmajorgrids,
		grid style={line width=.5pt, draw=gray!90,dashed},
		major grid style={line width=.2pt,draw=gray!50},
		bar width=7pt,
		ymin=0,
		ymax = 2500,
		legend columns=3,
		scaled y ticks=base 10:-3,
		]
		\addplot  [blue,fill=white,pattern=north east lines,pattern color=.,]coordinates {
			(0, 255) (1, 255) (2, 255) (3, 255)(4, 255)
		};
		\addplot coordinates {
			(0, 142) (1, 142) (2, 142) (3, 142)(4, 142)
		};
		\addplot  [black,fill=white,pattern=dots,pattern color=.,] coordinates {
			(0, 687) (1, 911) (2, 1319) (3, 1575)(4, 1612)
		};
		\legend{DWT, PCA, SVM}
	\end{axis}
\end{tikzpicture}\vspace{-.5em}
			\end{subfigure}
		}\vspace{-.9em}
		\caption{UCR-ECG}\label{fig:exp:detail:IM} 
	\end{subfigure}\\
	\begin{subfigure}{1\textwidth}
		\resizebox{.99\textwidth}{!}{
			\begin{subfigure}{.31\textwidth}
				\definecolor{A}{HTML}{e6194B}%
\definecolor{B}{HTML}{f58231}%
\definecolor{C}{HTML}{4363d8}%
\definecolor{D}{HTML}{911eb4}%
\definecolor{E}{HTML}{3cb44b}%
\definecolor{F}{rgb}{0.92900,0.69400,0.12500}%
\definecolor{G}{HTML}{808000}%
\definecolor{H}{HTML}{000000}%

\begin{tikzpicture} 
	\begin{axis}[
		ybar stacked,
		xtick={0,1,2,3,4,5},
		xticklabels = {$4$, $8$, $16$, $32$, $64$, $100$},
		width=.8\textwidth,
		height=.4\textwidth,
		at={(1.128in,0.894in)},
		ymin=0,
		scale only axis,
		xlabel={\# classes ($ s $)},
		ylabel = {\Prover~time (sec)},
		ylabel shift=-7pt,
		yticklabel shift={0cm},
		axis background/.style={fill=white},
		legend columns=3,
		legend style={legend cell align=left, align=left, fill=none, draw=none,inner sep=-0pt, row sep=0pt, font=\tiny},
		legend pos = north west,
		ymajorgrids,
		xmajorgrids,
		grid style={line width=.5pt, draw=gray!90,dashed},
		major grid style={line width=.2pt,draw=gray!50},
		bar width=7pt,
		legend columns=3,
		scaled y ticks=base 10:-3,
		]
		\addplot  [blue,fill=white,pattern=north east lines,pattern color=.,]coordinates {
			(0, 656) (1, 657) (2, 655) (3, 657)(4, 657)(5, 657)
		};
		\addplot coordinates {
			(0, 72) (1, 72) (2, 73) (3, 75)(4, 74)(5, 75)
		};
		\addplot  [black,fill=white,pattern=dots,pattern color=.,] coordinates {
			(0, 37) (1, 85) (2, 187) (3, 202)(4, 442)(5, 716)
		};
	\legend{DWT, PCA, SVM}
	\end{axis}
\end{tikzpicture}\vspace{-.5em}
			\end{subfigure}\hspace{4mm}
			\begin{subfigure}{.31\textwidth}
				\definecolor{A}{HTML}{e6194B}%
\definecolor{B}{HTML}{f58231}%
\definecolor{C}{HTML}{4363d8}%
\definecolor{D}{HTML}{911eb4}%
\definecolor{E}{HTML}{3cb44b}%
\definecolor{F}{rgb}{0.92900,0.69400,0.12500}%
\definecolor{G}{HTML}{808000}%
\definecolor{H}{HTML}{000000}%

\begin{tikzpicture} 
	\begin{axis}[
		ybar stacked,
		xtick={0,1,2,3,4,5},
		xticklabels = {$4$, $8$, $16$, $32$, $64$, $100$},
		width=.8\textwidth,
		height=.4\textwidth,
		at={(1.128in,0.894in)},
		scale only axis,
		xlabel={\# classes ($ s $)},
		ylabel = {\Verifier~time (sec)},
		ylabel shift=-7pt,
		yticklabel shift={0cm},
		axis background/.style={fill=white},
		legend columns=3,
		legend style={legend cell align=left, align=left, fill=none, draw=none,inner sep=-0pt, row sep=0pt,font=\tiny},
		legend pos = north west,
		ymajorgrids,
		xmajorgrids,
		grid style={line width=.5pt, draw=gray!90,dashed},
		major grid style={line width=.2pt,draw=gray!50},
		bar width=7pt,
		ymin=0,
		legend columns=3,
		]
		\addplot  [blue,fill=white,pattern=north east lines,pattern color=.,]coordinates {
			(0, 1.931) (1, 1.931) (2, 1.943) (3, 1.945)(4, 1.938)(5, 1.950)
		};
		\addplot coordinates {
			(0, .812) (1, .816) (2, .821) (3, .820)(4, .819)(5, .824)
		};
		\addplot  [black,fill=white,pattern=dots,pattern color=.,] coordinates {
			(0, .432) (1, .653) (2, .969) (3, 1.009)(4, 1.494)(5, 1.895)
		};
		\legend{DWT, PCA, SVM}
	\end{axis}
\end{tikzpicture}\vspace{-.5em}
			\end{subfigure}
			\begin{subfigure}{.31\textwidth}
				\definecolor{A}{HTML}{e6194B}%
\definecolor{B}{HTML}{f58231}%
\definecolor{C}{HTML}{4363d8}%
\definecolor{D}{HTML}{911eb4}%
\definecolor{E}{HTML}{3cb44b}%
\definecolor{F}{rgb}{0.92900,0.69400,0.12500}%
\definecolor{G}{HTML}{808000}%
\definecolor{H}{HTML}{000000}%

\begin{tikzpicture} 
	\begin{axis}[
		ybar stacked,
		xtick={0,1,2,3,4,5},
		xticklabels = {$4$, $8$, $16$, $32$, $64$, $100$},
		width=.8\textwidth,
		height=.4\textwidth,
		at={(1.128in,0.894in)},
		scale only axis,
		xlabel={\# classes ($ s $)},
		ylabel = {Proof size (KB)},
		ylabel shift=-7pt,
		yticklabel shift={0cm},
		axis background/.style={fill=white},
		legend columns=3,
		legend style={legend cell align=left, align=left, fill=none, draw=none,inner sep=-0pt, row sep=0pt,font=\tiny},
		legend pos = north west,
		ymajorgrids,
		xmajorgrids,
		grid style={line width=.5pt, draw=gray!90,dashed},
		major grid style={line width=.2pt,draw=gray!50},
		bar width=7pt,
		ymin=0,
		legend columns=3,
		scaled y ticks=base 10:-3,
		]
		\addplot  [blue,fill=white,pattern=north east lines,pattern color=.,] coordinates {
			(0, 1046) (1, 1046) (2, 1047) (3, 1046)(4, 1046)(5, 1046)
		};
		\addplot coordinates {
			(0, 579) (1, 579) (2, 580) (3, 578)(4, 579)(5, 579)
		};
		\addplot  [black,fill=white,pattern=dots,pattern color=.,] coordinates {
			(0, 179) (1, 270) (2, 401) (3, 417)(4, 618)(5, 785)
		};
			\legend{DWT, PCA, SVM}
	\end{axis}
\end{tikzpicture}\vspace{-.5em}
			\end{subfigure}
		}\vspace{-.7em}
		\caption{{Reduced Cifar-100}}
		\label{fig:exp:detail:cifar}
	\end{subfigure}\\
	
	\begin{subfigure}{1\textwidth}
		\resizebox{.99\textwidth}{!}{
			\centering
			\begin{subfigure}{.31\textwidth}
				\definecolor{A}{HTML}{e6194B}%
\definecolor{B}{HTML}{f58231}%
\definecolor{C}{HTML}{4363d8}%
\definecolor{D}{HTML}{911eb4}%
\definecolor{E}{HTML}{3cb44b}%
\definecolor{F}{rgb}{0.92900,0.69400,0.12500}%
\definecolor{G}{HTML}{808000}%
\definecolor{H}{HTML}{000000}%

%


\begin{tikzpicture} 
	\begin{axis}[
		ybar stacked,
		xtick={0,1,2,3,4,5,6, 7, 8},
		xticklabels = {$2^3$, $2^4$, $2^5$, $2^6$, $2^7$, $2^8$, $2^9$, $2^{10}$, $2^{11}$},
		width=.8\textwidth,
		height=.4\textwidth,
		at={(1.128in,0.894in)},
		scale only axis,
		xlabel={\# classes ($ s $)},
		ylabel = {\Prover~time (sec)},
		ylabel shift=-7pt,
		yticklabel shift={0cm},
		axis background/.style={fill=white},
		legend columns=3,
		legend style={legend cell align=left, align=left, fill=none, draw=none,inner sep=-0pt, row sep=0pt,font=\tiny},
		legend pos = north west,
		ymajorgrids,
		xmajorgrids,
		grid style={line width=.5pt, draw=gray!90,dashed},
		major grid style={line width=.2pt,draw=gray!50},
		bar width=7pt,
		legend columns=3,
		scaled y ticks=base 10:-3,
		ymin=0,
		]
		\addplot  [blue,fill=white,pattern=north east lines,pattern color=.,] coordinates {
			(0, 898) (1, 898) (2, 1209) (3, 1208)(4, 1209)(5, 1209)(6, 1208)(7, 1209)(8, 1209)
		};
		\addplot coordinates {
			(0, 99) (1, 99) (2, 134) (3, 133)(4, 133)(5, 135)(6, 133)(7, 134)(8, 135)
		};
		\addplot  [black,fill=white,pattern=dots,pattern color=.,] coordinates {
			(0, 704) (1, 1551) (2, 1704) (3, 2063)(4, 2064)(5, 3548)(6, 3908)(7, 4326)(8, 5628)
		};
		\legend{DWT, PCA, SVM}
	\end{axis}
\end{tikzpicture}\vspace{-.5em}
			\end{subfigure}\hspace{4mm}
			\begin{subfigure}{.31\textwidth}
				\definecolor{A}{HTML}{e6194B}%
\definecolor{B}{HTML}{f58231}%
\definecolor{C}{HTML}{4363d8}%
\definecolor{D}{HTML}{911eb4}%
\definecolor{E}{HTML}{3cb44b}%
\definecolor{F}{rgb}{0.92900,0.69400,0.12500}%
\definecolor{G}{HTML}{808000}%
\definecolor{H}{HTML}{000000}%

\begin{tikzpicture} 
	\begin{axis}[
		ybar stacked,
		xtick={0,1,2,3,4,5,6, 7, 8},
		xticklabels = {$2^3$, $2^4$, $2^5$, $2^6$, $2^7$, $2^8$, $2^9$, $2^{10}$, $2^{11}$},
		width=.8\textwidth,
		height=.4\textwidth,
		at={(1.128in,0.894in)},
		scale only axis,
		xlabel={\# classes ($ s $)},
		ylabel = {\Verifier~time (sec)},
		ylabel shift=-7pt,
		yticklabel shift={0cm},
		axis background/.style={fill=white},
		legend columns=3,
		legend style={legend cell align=left, align=left, fill=none, draw=none,inner sep=-0pt, row sep=0pt,font=\tiny},
		legend pos = north west,
		ymajorgrids,
		xmajorgrids,
		grid style={line width=.5pt, draw=gray!90,dashed},
		major grid style={line width=.2pt,draw=gray!50},
		bar width=7pt,
		ymin=0,
		legend columns=3,
		]
		\addplot  [blue,fill=white,pattern=north east lines,pattern color=.,] coordinates {
			(0, 2.257) (1, 2.257) (2, 2.618) (3, 2.618)(4, 2.619)(5, 2.617)(6, 2.618)(7, 2.619)(8, 2.618)
		};
		\addplot coordinates {
			(0, .950) (1, .951) (2, 1.102) (3, 1.123)(4, 1.124)(5, 1.118)(6, 1.130)(7, 1.126)(8, 1.119)
		};
		\addplot  [black,fill=white,pattern=dots,pattern color=.,] coordinates {
			(0, 1.885) (1, 2.797) (2, 2.935) (3, 3.230)(4, 3.229)(5, 4.235)(6, 3.834)(7, 4.677)(8, 5.335)
		};
		\legend{DWT, PCA, SVM}
	\end{axis}
\end{tikzpicture}\vspace{-.5em}
			\end{subfigure}
			\begin{subfigure}{.31\textwidth}
				\definecolor{A}{HTML}{e6194B}%
\definecolor{B}{HTML}{f58231}%
\definecolor{C}{HTML}{4363d8}%
\definecolor{D}{HTML}{911eb4}%
\definecolor{E}{HTML}{3cb44b}%
\definecolor{F}{rgb}{0.92900,0.69400,0.12500}%
\definecolor{G}{HTML}{808000}%
\definecolor{H}{HTML}{000000}%

\begin{tikzpicture} 
	\begin{axis}[
	ybar stacked,
	xtick={0,1,2,3,4,5,6, 7, 8},
	xticklabels = {$2^3$, $2^4$, $2^5$, $2^6$, $2^7$, $2^8$, $2^9$, $2^{10}$, $2^{11}$},
	width=.8\textwidth,
	height=.4\textwidth,
	at={(1.128in,0.894in)},
	scale only axis,
		xlabel={\# classes ($ s $)},
	ylabel = {Proof size (KB)},
	ylabel shift=-7pt,
	yticklabel shift={0cm},
	axis background/.style={fill=white},
	legend columns=3,
	legend style={legend cell align=left, align=left, fill=none, draw=none,inner sep=-0pt, row sep=0pt,font=\tiny},
	legend pos = north west,
	ymajorgrids,
	xmajorgrids,
	grid style={line width=.5pt, draw=gray!90,dashed},
	major grid style={line width=.2pt,draw=gray!50},
	bar width=7pt,
	ymin=0,
	legend columns=3,
	scaled y ticks=base 10:-3,
		]
		\addplot  [blue,fill=white,pattern=north east lines,pattern color=.,] coordinates {
			(0, 1223) (1, 1224) (2, 1420) (3, 1419)(4, 1419)(5, 1420)(6, 1421)(7, 1421)(8, 1421)
		};
		\addplot coordinates {
			(0, 676) (1, 676) (2, 785) (3, 784)(4, 784)(5, 784)(6, 784)(7, 785)(8, 784)
		};
		\addplot  [black,fill=white,pattern=dots,pattern color=.,] coordinates {
			(0, 779) (1, 1157) (2, 1214) (3, 1334)(4, 1335)(5, 1751)(6, 1584)(7, 1933)(8, 2206)
		};
		\legend{DWT, PCA, SVM}
	\end{axis}
\end{tikzpicture}\vspace{-.5em}
			\end{subfigure}
		}\vspace{-.5em}
		\caption{{Reduced LFW}}
		\label{fig:exp:detail:LFW} 
	\end{subfigure}\vspace{-.7em}
	\caption{Detailed cost of \sys.}
	\label{fig:exp:detail}
\end{figure*}

\subsection{Overall Results}\label{sec:exp:overall}
{\sys~is one to three orders of magnitudes more efficient than the baseline in \emph{all} metrics.}
\autoref{fig:exp:overall} presents the performance of our technique 
compared with the baseline approach in terms of proving time, verification time, and proof size, in three datasets with different sizes.
For example, on UCR-ECG dataset, our proving time is from 321 to 518 seconds for $4$ to $42$ classes, 
while it takes from 1429 to 2807 seconds if using the baseline approach.
The gap between our scheme and the baseline is more significant when the number of classes increases.
Specifically, on LFW dataset, with 8 classes,
our scheme achieves $6.75\times$ faster proving time, where it only takes 1702 seconds, compared with 11491 seconds in the baseline.
With 2048 classes, our proving time is 6977 seconds, approximately 1842$\times$ faster than the baseline, which takes 2439811 seconds. The verification time and proof size follow a similar trend, 
in which \sys~achieves an order of magnitude faster verification time and smaller proof size than the baseline.
Specifically, on LFW dataset, the verification time is 6.6 seconds for 16 classes and 19.2 seconds in the baseline. 
The proof size is 3059 KB in our scheme, compared with 11946 KB in the baseline. 
On the LFW dataset with 2048 classes, our verification time is $9$ seconds, and the proof size is $4411$ KB, while it takes $ 123.6 $ seconds for verification with 56856 KB proof size in the baseline.
This results in around $12 \times$ faster on the verification time and $ 14 \times $ smaller proof size, respectively.

We can also see the verification and bandwidth in $ \sys $ are highly efficient, i.e., less than $ 10$ seconds and $ 5$ MB, respectively, compared with the proving.
This is because we use Spartan as the CP-ZKP backend, which offers sublinear verification and proof size overhead.

The concrete end-to-end computation latency and communication in \autoref{fig:exp:overall} also confirm the efficiency improvement of our optimization techniques.
By introducing the split technique and employing the random linear combination, the complexity is reduced from $O(mc+mk)$ to $O(c^2+m)$, where $ c $ is a very small constant in practice (e.g., $ c = 4 $ for {Daubechies DB4 DWT}). The most significant improvement in the overall cost is achieved when the number of classes is large. That is due to the employment of $\mathsf{Max}$ and $\mathsf{Exp}$ gadgets in the SVM phase, which reduces the complexity from $O(s^2)$ to $O(s)$.
Such asymptotic improvement helps to achieve one to three orders of magnitude faster computation time and lower communication overhead on real datasets.

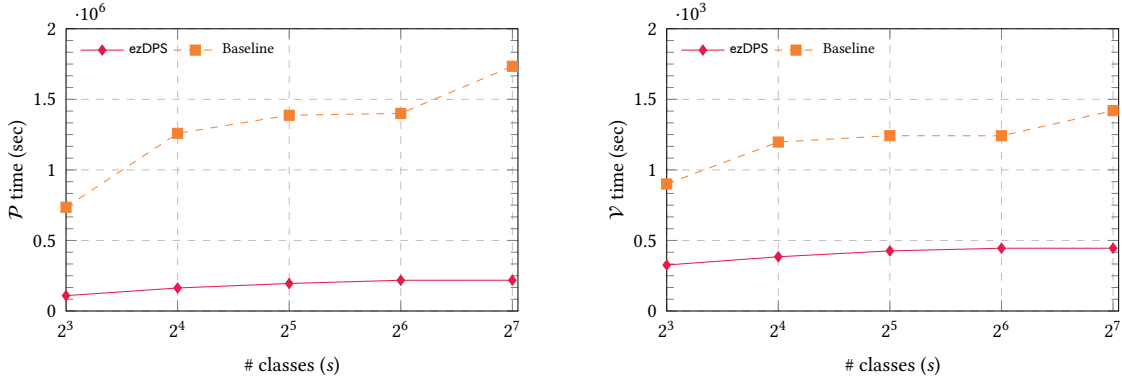
\begin{figure}[!t]
\centering
\captionsetup[subfigure]{justification=centering}
\begin{subfigure}{1\columnwidth}
\centering
	\resizebox{.98\columnwidth}{!}{
		\begin{subfigure}{.49\columnwidth}
%
%
\definecolor{A}{HTML}{e6194B}%
\definecolor{B}{HTML}{f58231}%
\definecolor{C}{HTML}{4363d8}%
\definecolor{D}{HTML}{911eb4}%
\definecolor{E}{HTML}{3cb44b}%
\definecolor{F}{rgb}{0.92900,0.69400,0.12500}%
\definecolor{G}{HTML}{808000}%
\definecolor{H}{HTML}{000000}%
\begin{tikzpicture}
	
	\begin{axis}[%
		width=.8\textwidth,
		height=.5\textwidth,
		at={(1.128in,0.894in)},
		scale only axis,
		xmin=0,
		xmax=4.05,
		xlabel={\# classes ($s$) },
		xtick={0,1,2,3,4},
		xticklabels={$2^3$, $ 2^4 $, $ 2^5 $, $ 2^6 $, $2^7$},
		ymin=0,
		ymax=2000000,
		ytick distance=500000,
		ylabel = {\Prover~time (sec)},
		ylabel shift=-5pt,
		yticklabel shift={0cm},
		axis background/.style={fill=white},
		legend columns=3,
		legend style={legend cell align=left, align=left, fill=none, draw=none,inner sep=-0pt, row sep=0pt, font=\tiny},
		legend pos = north west,
		ymajorgrids,
		xmajorgrids,
		grid style={line width=.5pt, draw=gray!90,dashed},
		major grid style={line width=.2pt,draw=gray!50},
		minor y tick num=5,
		]
		\addplot [color=A, solid, mark=diamond*, mark options={solid, A}]
		table[row sep=crcr]{%
			0 108928\\
			1 163136\\
			2 195072\\
			3 217920\\
			4 218048\\
		};
		\addlegendentry{\sys}
		
		\addplot [color=B, dashed, mark=square*, mark options={solid, B}]
		table[row sep=crcr]{%
			0 735424\\
			1 1259168\\
			2 1387264\\
			3 1400544\\
			4 1734080\\
		};
		\addlegendentry{Baseline}		
	\end{axis}
\end{tikzpicture}%
\vspace{-.5em}
		\end{subfigure}\hspace{4mm}
		\begin{subfigure}{.49\columnwidth}
%
%
\definecolor{A}{HTML}{e6194B}%
\definecolor{B}{HTML}{f58231}%
\definecolor{C}{HTML}{4363d8}%
\definecolor{D}{HTML}{911eb4}%
\definecolor{E}{HTML}{3cb44b}%
\definecolor{F}{rgb}{0.92900,0.69400,0.12500}%
\definecolor{G}{HTML}{808000}%
\definecolor{H}{HTML}{000000}%
\begin{tikzpicture}
	
	\begin{axis}[%
		width=.8\textwidth,
		height=.5\textwidth,
		at={(1.128in,0.894in)},
		scale only axis,
		xmin=0,
		xmax=4.05,
		xlabel={\# classes ($s$) },
		xtick={0,1,2,3, 4},
		xticklabels={$2^3$, $ 2^4 $, $ 2^5 $, $ 2^6 $, $2^7$},
		ymin=0,
		ymax=2000,
		ytick distance=500,
		ylabel = {\Verifier~time (sec)},
		ylabel shift=-5pt,
		yticklabel shift={0cm},
		axis background/.style={fill=white},
		legend columns=3,
		legend style={legend cell align=left, align=left, fill=none, draw=none,inner sep=-0pt, row sep=0pt, font=\tiny},
		legend pos = north west,
		ymajorgrids,
		xmajorgrids,
		grid style={line width=.5pt, draw=gray!90,dashed},
		major grid style={line width=.2pt,draw=gray!50},
		minor y tick num=5,
		scaled y ticks=base 10:-3,
		]
		\addplot [color=A, solid, mark=diamond*, mark options={solid, A}]
		table[row sep=crcr]{%
			0 326.016\\
			1 384.384\\
			2 426.048\\
			3 444.864\\
			4 444.8\\
		};
		\addlegendentry{\sys}
		
		\addplot [color=B, dashed, mark=square*, mark options={solid, B}]
		table[row sep=crcr]{%
			0 900.16\\
			1 1197.376\\
			2 1242.176\\
			3 1241.664\\
			4 1419.648\\
		};
		\addlegendentry{Baseline}		
	\end{axis}
\end{tikzpicture}%
\vspace{-.5em}
		\end{subfigure}
	}\vspace{-.5em}
\end{subfigure}\\
\caption{Performance of zkPoA on {reduced LFW}.}
\label{fig:exp:poa}
\end{figure}

{Finally, we report the performance of zkPoA scheme proposed in \autoref{sec:zkPoA}. 
	Since zkPoA is derived from the proof of inference for individual samples, 
	our scheme maintains the same ratio of performance gain over the baseline as reported in \autoref{sec:exp:overall}.
	Concretely, we tested zkPoA on the reduced LFW dataset with 64 samples. 
	As shown in \autoref{fig:exp:poa}, we achieve $6\times$ to $9\times$ faster on the prover's time, $3\times$ faster on the verifier's time compared with the baseline. 
	Regarding proof size, our scheme incurs 171392--226432 KB, which is three times smaller than the baseline that requires 576148--827968 KB.
	The complexity of zkPoA is linear with the number of samples, and 
	its main overhead stems from the inference proof of individual samples.
    
    We can see that our zkPoA scheme currently only supports plain accuracy verification, meaning the proof is given only for a specific test set. 
    In the ML setting, cross-validation over different test sets is generally applied to report a more reliable accuracy result. 
    It is interesting to explore if an zkPoA scheme can permit accuracy verification with cross-validation without leaking the model privacy due to multiple test sets. We leave it as an open research problem for future investigation.
}

\begin{table*}[t]
	\centering
	\small
	\caption{{Inference accuracy of ML algorithms on whole datasets.}}
	\label{tab:acc}
	\resizebox{1\textwidth}{!}{
	\begin{threeparttable}
	\begin{tabular}{|l|c|c|c|c||c|c|c|c|c||c|c|c|c|c|}
		\hline
		\textbf{Method} & 
		\multicolumn{4}{c||}{\textbf{UCR-ECG}} & \multicolumn{5}{c||}{\textbf{Cifar-100}} & \multicolumn{5}{c|}{\textbf{LFW}} \\ \cline{1-15} 
		\# classes &  $ 8 $ &  $ 16 $ & $ 32 $ & $ 42 $  & $ 8 $ & $ 16 $ & $ 32 $ & $ 64 $ & $100$ & $ 8 $  & $ 16 $ & $ 32 $ & $ 64 $  & $128$ \\ \hline\hline
		DT only & 0.84$\pm$0.09 & 0.73$\pm$0.10 & 0.65$\pm$0.06 & 0.65$\pm$0.01 & 0.37 & 0.23 & 0.17 & 0.11 & 0.09 &0.47$\pm$0.06 & 0.36$\pm$0.08 & 0.27$\pm$0.09 & 0.21$\pm$0.05 & 0.10$\pm$0.03\\ \hline
		DWT+PCA+DT & 0.79$\pm$0.07 & 0.77$\pm$0.07 & 0.65$\pm$0.07 & 0.66$\pm$0.04 & 0.32& 0.22 & 0.17 & 0.11 & 0.09 & 0.43$\pm$0.07 & 0.32$\pm$0.05 & 0.24$\pm$0.05 & 0.17$\pm$0.07 & 0.15$\pm$0.04\\ \hline\hline
		SVM only  & 0.96$\pm$0.01 & 0.96$\pm$0.01 & 0.91$\pm$0.04& 0.91$\pm$0.02& 0.13 & 0.07 & 0.03 & 0.02 &0.01 & 0.39$\pm$0.07 & 0.30$\pm$0.07 & 0.23$\pm$0.06 & 0.18$\pm$0.07 & 0.08$\pm$0.2\\  \hline\hline
		DWT+PCA+SVM & 0.99$\pm$0.03 & 0.97$\pm$0.04 & 0.93$\pm$0.03 & 0.92$\pm$0.05 & 0.55 & 0.41 & 0.35 & 0.29 & 0.24 & 0.73$\pm$0.07 & 0.60$\pm$0.08 & 0.48$\pm$0.06 & 0.36$\pm$0.07 & 0.20$\pm$0.06\\ \hline

		DWT+PCA+SVM (FPA)\textsuperscript{$\ddagger$}  & 0.97$\pm$0.03 & 0.95$\pm$0.04 & 0.91$\pm$0.02& 0.91$\pm$0.06& 0.55 & 0.4 & 0.35& 0.28 & 0.24 & 0.73$\pm$0.07 & 0.6$\pm$0.06 & 0.47$\pm$0.06 & 0.36$\pm$0.06 & 0.19$\pm$0.05 \\ 
		\hline
	\end{tabular}
	\begin{tablenotes}[flushleft]
	    \item $\ddagger$ FPA stands for fixed-point arithmetic.
	\end{tablenotes}
	\end{threeparttable}
	}
\end{table*}

\subsection{Detailed Cost Analysis}
\label{sec:exp:detail_cost}

We dissected the total cost of our scheme to investigate the impact of each data processing on the overall performance.
\autoref{fig:exp:detail} presents the detailed cost of \sys~with three datasets.
In \sys, the sample was processed in three phases, including DWT noise reduction, PCA feature extraction, and SVM classification.

\noindent $ \bullet $\underline{\textit{DWT Processing:}}
{The cost of DWT processing is stable when varying the number of classes ($ s $) and contributes a considerable portion to the overall performance.}
This is because the complexity of DWT is independent of $ s $, i.e., $ O(mn) $, which is bigger than PCA (i.e., $ O(m) $), but smaller than  SVM (i.e., $O((n+k)t + ns)$) for a large number of classes.
On the UCR-ECG dataset, the proving time is around 160 seconds, and the verification time and proof size are around 0.47 seconds and 256 KB, respectively. 
On Cifar-100, the proving time, verification time, and proof size are around 656 seconds, 1.94 seconds, and 1046 KB, respectively. On LFW dataset, the performance of the DWT phase ranges from 898 to 1209 seconds, 2.2 to 2.6 seconds, and 676 to 1421 KB, respectively.
There is a considerable difference in proving DWT across three datasets. That is because the dimension of inputs varies between different datasets, e.g., $m$ equals 750, 3072, and 4200 (or 5655) on UCR-ECG, Cifar-100, and LFW datasets, respectively.

\noindent $ \bullet $\underline{\textit{PCA-based Feature Extraction:}}
{The cost of PCA processing is stable even when the number of classes $ s $ increases and it contributes the least portion to the overall performance of our scheme.}
This is because the complexity of PCA is $ O(m) $ (which is also independent to $ s $), compared with $O(nm)$ in DWT and $O((n+k)t + ns)$ in SVM.
For example, it costs around 17 seconds for proving, 0.198 seconds for the verification, and around 141 KB for the proof size on the UCR-ECG dataset. 
The cost of proving PCA is nearly negligible on UCR-ECG and LFW datasets.
This is because the number of constraints for PCA is relatively small (i.e., 750 on UCR-ECG and 5655 on LFW) compared with DWT and SVM (e.g., on UCR-ECG, there are 75439 and over 110322 constraints in DWT and SVM, respectively).
Since the verification time and proof size is sublinear, the proportion of PCA processing becomes larger relatively compared with DWT and SVM.

\noindent $ \bullet $\underline{\textit{SVM Classifcation:}}
{SVM computation is the most dominant factor, especially on large datasets (with more than 256 classes), which contributes over 73\% to the total proving cost.}
That is because the cost of SVM is $O((n+k)t + ns)$, and thus it grows linearly with $s$.
Notice that the increase of the number of classes also leads to the increase of the model parameters ($ t $).
On UCR-ECG dataset, the proving time of SVM ranges from 142 to 339 seconds. The verification time is from 1.65 to 4.24 seconds, and the proof size is 688 KB to 1623 KB for $4$ to $42$ classes. 
On Cifar-100 dataset, the proving time of SVM costs from 43 to 722 seconds, while its verification time and proof size are from 0.4 to 1.925 seconds and 179 KB to 785 KB, respectively, for $4$ to $100$ classes.
On the LFW dataset, the proving time ranges from 704 to 6011 seconds for 8 to 2048 classes, while the verification time ranges from 1.89 to 5.73 seconds, and the proof size ranges from 779 to 2286 KB, respectively. 
The gap between SVM vs. DWT and PCA looks smaller in the verification time and proof size due to their sublinear growth of complexity by Spartan ZKP.

\PP{Estimated performance on whole datasets}
{
Based on the overall results (\autoref{sec:exp:overall}) and the above cost analysis on the reduced datasets, 
we estimated the cryptographic overhead of our scheme when tested on the whole Cifar-100 and LFW. 
For $X \in \{8,16,32,64,100\} $ classes in Cifar-100 with the standard train/test method, 
the proving time of our scheme is estimated to take 8189 to 108698 seconds. 
The verification time and proof size are estimated to take 8.8-26 seconds  and 4154--11247 KB, respectively. 
In LFW dataset with $X$ most sampled classes, 
the proving time is estimated to take 5823 to 24772 seconds, 
while the verification time and proof size is estimated to take 9.24--16.34 seconds and 4487--7424 KB, respectively.
The estimated proving time is significant, since the estimation is based on our current hardware (i.e., a laptop without multi-threading).
In practice, since the prover is the server that generally has better computational resource (e.g., multi-core CPU with higher frequency and multi-threading), we expect the actual proving time will be significantly faster.
For the whole Cifar-100, since the number of support vectors ($t$) is large, it incurs a large model size, resulting in high proving time.
We expect that once some optimization techniques (e.g., \cite{svm:optimization, svm:optimization2}) are applied to reduce the model complexity, all the cryptographic overhead will be significantly reduced.
We leave such optimization as our future work.}

\subsection{Accuracy}\label{sec:exp:accuracyloss}

{
We report the accuracy of ML algorithms on the whole dataset of UCR-ECG, Cifar-100, and LFW. 
In Cifar-100, we used all data from classes  $0,1,\dots,X-1$ for $X \le 100 $ classes and tested with its standard train/test method. 
For LFW and UCR-ECG, since there is no standard train/test split, 
we applied the cross-validation to report the accuracy. 
In LFW, since the number of samples in each class is unbalanced,
we selected $X$ classes that have the most data samples.
In UCR-ECG, we chose data from classes $0,1,\dots,X-1$. 
\autoref{tab:acc} presents the plain accuracy of ML algorithms on the selected datasets.
The last row of \autoref{tab:acc} presents the accuracy of executing DWT+PCA+SVM inference with Fixed-Point Arithmetic (FPA), which is similar to how our \sys~works. 
We can see that FPA leads to an accuracy decrease of around 1\% to 2\%. In LFW, DWT+PCA+SVM with floating-point arithmetic achieves $73\%\pm7\%$ and $60\%\pm8\%$ accuracy rates for $8$ and $16$ classes, respectively. The accuracy decreases $1\%$ to $2\%$, leading to the accuracy rates of $72\%\pm7\%$ and $60\%\pm6\%$, respectively. A similar trend is also observed in UCR-ECG and Cifar-100 datasets, where the accuracy loses around 1\% to 2\% due to FPA.}

%
%
For curious readers, we conservatively report the best inference accuracy that each of our benchmark datasets currently achieves with different state-of-the-art ML pipeline techniques (without integrity and model privacy). UCR-ECG can achieve 97.5\% accuracy by combining Gated Recurrent Unit with Fully Convolutional Network \cite{acc:ucr:elsayed2018deep}. Cifar-100 can achieve 96.08\% accuracy by combining ImageNet pre-trained model with sharpness-aware minimization \cite{acc:cifar100:foret2020sharpness}. Finally, LFW can achieve 99\% accuracy using optimized VarGNet \cite{acc:lfw:yan2019vargfacenet}. Since these pipeline techniques are highly optimized for each dataset, they yield higher accuracy than our generic framework. We leave the investigation on zero-knowledge proofs for optimization techniques that can be integrated into our framework to further improve the accuracy of our future work.



\section{Related Work}\label{sec:relatedwork}

\PP{Privacy-Preserving ML}
Privacy-Preserving ML (PPML) permits secure evaluation of ML computation 
without leaking information about the ML model and training/testing data.
Most PPML techniques rely on either secure computation protocols such as Multi-party computation (MPC) \cite{MPC:cramer2015secure} and Homomorphic Encryption (HE) \cite{HE:gentry2009fully}, or Trusted Execution Environment (TEE) such as Intel-SGX \cite{TEE:SGX:costan2016intel}.
PPML has been investigated in both training and inference phases.
Many PPML training schemes have been proposed for established ML algorithms such as 
decision tree \cite{agrawal2000privacy}, k-means clustering \cite{jagannathan2005privacy, bunn2007secure}, SVM \cite{vaidya2008privacy}, linear regression \cite{mohassel2018aby3,mohassel2017secureml}, logistic regression (LR) \cite{272298, mohassel2018aby3} and neural networks (NN) \cite{mohassel2018aby3}. 
Other frameworks focus on the inference phase such as GAZELLE \cite{juvekar2018gazelle}, SWIFT \cite{272298}, MiniONN \cite{2017Oblivious}, XONN \cite{2019XONN}, CHET \cite{dathathri2019chet}, Delphi \cite{mishra2020delphi}, CryptoNets \cite{gilad2016cryptonets} and its variants \cite{PPML:cryptodl, brutzkus2019low}. 
Given MPC and FHE incur high costs in large-scale data processing, 
some studies harnessed Intel-SGX to make PPML more practical \cite{PPML:SGX:ohrimenko2016oblivious}. 
Unlike our \sys~or zkML, PPML protects the privacy of client and server data but not computation integrity.

\PP{Verifiable and zero-knowledge ML}
Unlike PPML, verifiable ML (vML) and zkML focus on the integrity of delegated ML computation using VC and zero-knowledge techniques \cite{setty2020spartan, 2016Quadratic, parno2013pinocchio, goldwasser2015delegating,zkp:legosnark}. 
Both vML and zkML are still in the early development stage, with a limited number of schemes being proposed.
In vML, the resource-limited client delegates the training/inference tasks to the server, 
and later checks if the task has been performed correctly (no privacy guarantee).
Zhao et al. \cite{zhao2021veriml} proposed VeriML, a vML framework for linear regression, LR, NN, SVM, and DT training. Some vML schemes are designed for DNN inference (e.g., \cite{ghodsi2017safetynets,tramer2018slalom}) using VC protocols (e.g., \cite{groth2016size, goldwasser2015delegating}) or TEE \cite{TEE:SGX:costan2016intel}. 
On the other hand, zkML, first studied in 2020 \cite{zhang2020zero}, 
enables integrity and model privacy in the inference phase, 
where the client can verify if the inference result on her data is indeed computed from the server's committed model without learning the model parameters. 
Zhang et al. designed a zkDT scheme \cite{zhang2020zero}, followed by a few zero-knowledge DNN inference constructions \cite{lee2020vcnn, feng2021zen,liu2021zkcnn}.
Weng et al. proposed Mystique \cite{274713}, a zkVC compiler for efficient zero-knowledge NN inference.

\section{Conclusion}
We proposed \sys, an efficient and zero-knowledge MLIP instantiated with effective ML algorithms including DWT, PCA, and SVM. We introduced new gadgets for proving ML operations in arithmetic circuits more effectively than generic approaches. We fully implemented our \sys~and evaluated its performance on real-world datasets. Experimental results showed that \sys~is highly efficient, which achieves orders of magnitudes more efficient than generic approaches.

\section*{Acknowledgement}
We would like to thank our shepherd and the anonymous reviewers in PETS 2023 for their insightful comments and suggestions to improve the quality of this paper.
Thang Hoang is supported by an unrestricted gift from Robert Bosch, and the Commonwealth Cyber Initiative (CCI), an investment in the advancement of cyber R\&D, innovation, and workforce development. For more information about CCI, visit www.cyberinitiative.org.
Haodi Wang is sponsored by the National Natural Science Foundation of China under Grants 62177007, 62102035, 61571049, 71961022, the Fundamental Research Funds for the Central Universities under Grants 2020NTST32.

\bibliographystyle{abbrv}
 
\bibliography{ref}
\appendix
\section{Example of Split Technique and Application}\label{sec:example:split}
We present a concrete example to demonstrate how the split technique reduces the number of constraints in proving DWT.
Suppose the input data is $ \vect{x} = [x_1, x_2, x_3,x_4,x_5,x_6]$, the low-pass filter  
is $ \vect{h} = [h_1, h_2, h_3, h_4]$.
Directly computing the first half of the DWT frequency component $ \vect{y} = [y_1, y_2, y_3] $, where 
\begin{equation}\label{eq:eg_ori}
    \begin{aligned}
        y_1 = x_1h_1+x_2h_2+x_3h_3+x_4h_4 \\
        y_2 = x_3h_1+x_4h_2+x_5h_3+x_6h_4 \\
        y_3 = x_5h_1+x_6h_2+x_1h_3+x_2h_4 \\
    \end{aligned}
\end{equation}
requires $12$ multiplications. The above computation can be combined by adopting the random linear combination, such that

\begin{equation}\label{eq:eg_spl}
    \begin{aligned}
    &(\alpha^3 h_1 + \alpha^2 h_2 + \alpha h_3 + h_4) (x_1 + \alpha x_2 + \alpha^2 x_3 +...+\alpha^5 x_6)\\
    &= \alpha^3 y_1 + \alpha^5 y_2 + y_3 - D
    \end{aligned}
\end{equation}
where $ D $ is the terms that have to be subtracted from the left side of \autoref{eq:eg_spl} such that

\begin{equation}\label{eq:split_and_dwt_appen}
    \begin{aligned}
    D &= x_1 h_4+\alpha(x_1h_3 + x_2 h_4)\\
    & + \alpha^2(x_1h_2 + x_2h_3 + x_3h_4) \\
    & + \alpha^4 (x_2h_1 + x_3h_2 + x_4h_3 + x_5h_4)\\
    & + \alpha^6 (x_4h_1 + x_5h_2 + x_6h_3)\\
    & + \alpha^7 (x_5h_1 + x_6h_2) + \alpha^8 x_6h_1 - y_3
    \end{aligned}
\end{equation}
We can see that there are 20 multiplications in $ D $ as the step of the sliding window between two rounds is two (i.e., computing $y_1$ starts with $x_{1}$,  while computing $y_2$ starts with $x_{3}$). 

\begin{figure*}[t!]
	\centering
	\includegraphics[width=\textwidth]{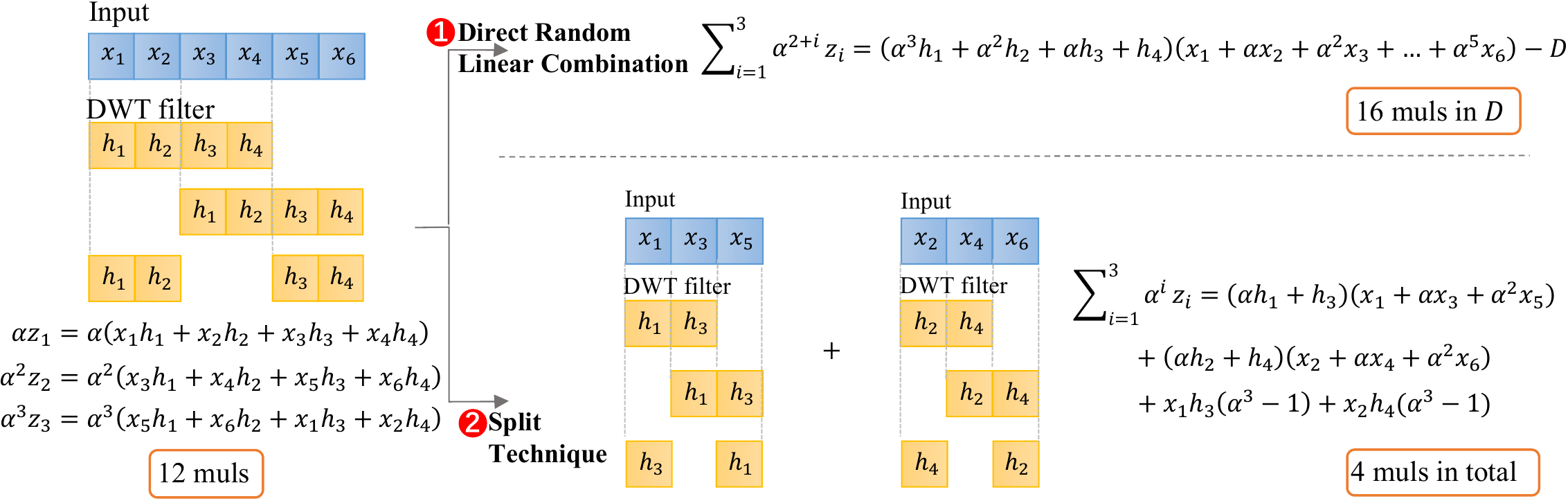}
	\caption{{Example of split technique applied to DWT decomposition vs. directly using random linear combination.}}
	\label{fig:dwt}
\end{figure*}

To improve the efficiency, the splitting technique separates the data sample and the low-pass filter into two parts, 
i.e., the odd part and the even part. 
Specifically, let $ \vect{x}^{(1)} = [x_1, x_3, x_5]$, $ \vect{x}^{(2)} = [x_2, x_4, x_6]$, $\vect{h}^{(1)} = [h_1, h_3]$, and $\vect{h}^{(2)} = [h_2, h_4]$. Therefore, \autoref{eq:eg_ori} is equivalent to
\begin{equation*}
    \begin{aligned}
        &(x_1 + \alpha x_3 + \alpha^2 x_5) (\alpha h_1 + h_3) + (x_2 + \alpha x_4 + \alpha^2 x_6) (\alpha h_2 +h_4) \\
        &=\Sigma_{i=1}^3 \alpha^i y_i - {D'} \\
        &=\Sigma_{i=1}^3 \alpha^i y_i - (x_1 h_3 (\alpha^3-1) + x_2h_4 (\alpha^3 - 1))
    \end{aligned}
\end{equation*}
which only requires 4 multiplications to prove compared with 20 in {\autoref{eq:split_and_dwt_appen}}. It reduces the number of intermediate terms in {$D'$}, thereby reducing the number of constraints. {We present the above toy example in \autoref{fig:dwt}}.

\begin{figure*}[t!]
	\centering
	\includegraphics[width=\textwidth]{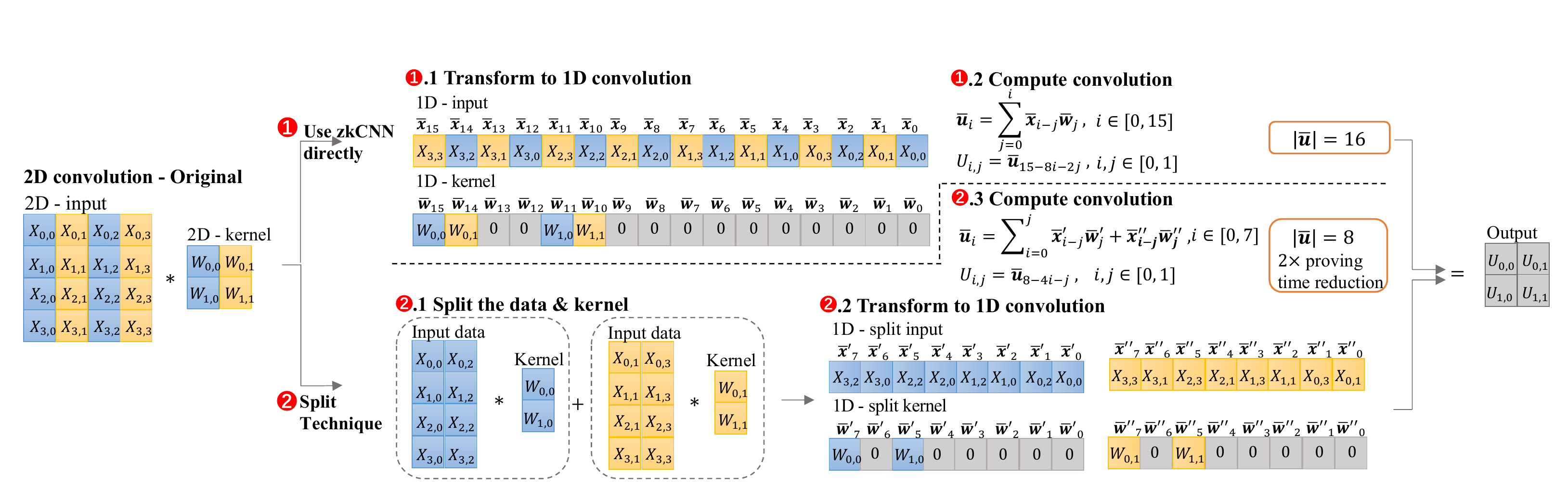}
	\caption{Adopting split technique to convolutions in zkCNN when sliding step $s=2$.}
	\label{fig:zkcnnexample}
\end{figure*}

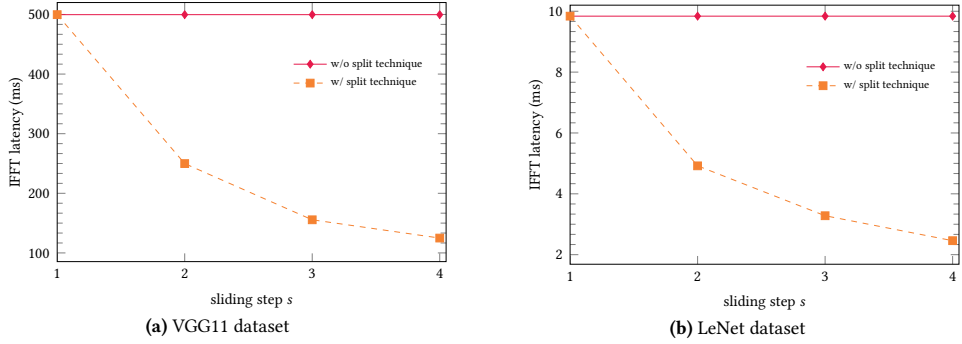
\begin{figure}[t]
	\centering
	\captionsetup[subfigure]{justification=centering}
	\begin{subfigure}{1\columnwidth}	
	\centering
		\resizebox{.8\columnwidth}{!}{
		\centering
			\begin{subfigure}{.49\columnwidth}
%
%
\definecolor{A}{HTML}{e6194B}%
\definecolor{B}{HTML}{f58231}%
\definecolor{C}{HTML}{4363d8}%
\definecolor{D}{HTML}{911eb4}%
\definecolor{E}{HTML}{3cb44b}%
\definecolor{F}{rgb}{0.92900,0.69400,0.12500}%
\definecolor{G}{HTML}{808000}%
\definecolor{H}{HTML}{000000}%
\begin{tikzpicture}
	
	\begin{axis}[%
		width=.9\textwidth,
		height=.6\textwidth,
		at={(1.128in,0.894in)},
		scale only axis,
		xmin=0,
		xmax=3.05,
		xlabel={sliding step $s$},
		xtick={0,1,2,3},
		xticklabels={$1$, $ 2 $, $ 3 $, $ 4 $},
		ymin=,
		ymax=520,
		ytick distance=100,
		ylabel = {IFFT latency (ms)},
		ylabel shift=-5pt,
		yticklabel shift={0cm},
		axis background/.style={fill=white},
		legend columns=1,
		legend style={legend cell align=left, align=left, fill=none, draw=none,inner sep=-0pt, row sep=0pt, font=\tiny, at={(.95,0.8)},anchor=north east},
		%
		minor y tick num=5,
		]
		\addplot [color=A, solid, mark=diamond*, mark options={solid, A}]
		table[row sep=crcr]{%
			0 499.670084\\
			1 499.670084\\
			2 499.670084\\
			3 499.670084\\
		};
		\addlegendentry{w/o split technique}
		
		\addplot [color=B, dashed, mark=square*, mark options={solid, B}]
		table[row sep=crcr]{%
		    0 499.670084\\
			1 249.835024\\
			2 155.556695\\
			3 124.917521\\
		};
		\addlegendentry{w/ split technique}		
	\end{axis}
\end{tikzpicture}%
\vspace{-.5em}
				\caption{VGG11 dataset }
			\end{subfigure}\hspace{15mm}
		
			\begin{subfigure}{.49\columnwidth}
%
%
\definecolor{A}{HTML}{e6194B}%
\definecolor{B}{HTML}{f58231}%
\definecolor{C}{HTML}{4363d8}%
\definecolor{D}{HTML}{911eb4}%
\definecolor{E}{HTML}{3cb44b}%
\definecolor{F}{rgb}{0.92900,0.69400,0.12500}%
\definecolor{G}{HTML}{808000}%
\definecolor{H}{HTML}{000000}%
\begin{tikzpicture}
	
	\begin{axis}[%
		width=.9\textwidth,
		height=.6\textwidth,
		at={(1.128in,0.894in)},
		scale only axis,
		xmin=0,
		xmax=3.05,
		xlabel={sliding step $s$ },
		xtick={0,1,2,3},
		xticklabels={$1$, $ 2 $, $ 3 $, $ 4 $},
		ymin=,
		ymax=10.2,
		ytick distance=2,
		ylabel = {IFFT latency (ms)},
		ylabel shift=-5pt,
		yticklabel shift={0cm},
		axis background/.style={fill=white},
		legend columns=1,
		legend style={legend cell align=left, align=left, fill=none, draw=none,inner sep=-0pt, row sep=0pt, font=\tiny,  at={(.95,0.8)}, anchor=north east},
		%
		minor y tick num=5,
		]
		\addplot [color=A, solid, mark=diamond*, mark options={solid, A}]
		table[row sep=crcr]{%
			0 9.838827\\
			1 9.838827\\
			2 9.838827\\
			3 9.838827\\
		};
		\addlegendentry{w/o split technique}
		
		\addplot [color=B, dashed, mark=square*, mark options={solid, B}]
		table[row sep=crcr]{%
			0 9.838827\\
			1 4.9194135\\
			2 3.279609\\
			3 2.459706\\
		};
		\addlegendentry{w/ split technique}		
	\end{axis}
\end{tikzpicture}%
\vspace{-.5em}
				\caption{LeNet dataset}
			\end{subfigure}
		}\vspace{-.5em}
	\end{subfigure}\\
	\caption{IFFT delay in zkCNN w/ or w/o split technique. }
	\label{fig:exp:zkcnn}
\end{figure}

\PP{Application to zkCNN}
We show that the split technique can be used to improve the efficiency of zkCNN \cite{liu2021zkcnn} in some cases when the sliding step $s$ between two rounds of convolution is larger than 1.
Note that $s \geq 2$ is generally adopted in deep learning regions \cite{krizhevsky2012imagenet}.

Suppose the input matrix $\vect{X}$ is of size $n \times n$ and the kernel matrix $\vect{W}$ is of size  $w \times w$.
The 2-D convolution between these two matrices is a matrix $\vect{U}$ of size $(\frac{n-w}{s}+1)\times (\frac{n-w}{s}+1)$ such that 

\begin{equation}\label{eq:zkcnn:conv}
	\vect{U}[i][j] = \sum_{u=0, v=0}^{w-1, w-1} \vect{X}[si+u][sj+v]\cdot \vect{W}[u][v]
\end{equation}
for $0 \le i,j \le (n/s-1)$. 

By zkCNN, the input and kernel matrices are first transformed to 1-D vectors to reduce the computation. Specifically, let $\bar{\vect{x}}, \bar{\vect{w}}, \bar{\vect{u}} \in \F^{n^2}$ be

\begin{equation} \label{eq:zkcnn:barXWU}
	\begin{aligned}
		\bar{\vect{x}}[un+v] &= \vect{X}[n-1-u][n-1-v] \text{, } 0\leq u <n, 0\leq v < n \\
		\bar{\vect{w}}[un+v] &= \begin{cases}
			\vect{W}[u][v], \text{ } 0\leq u,v < w \\
			0, \text{otherwise} 
		\end{cases}\\
		\bar{\vect{u}}[i] &= \sum_{j=0}^i \bar{\vect{x}}[i-j]\bar{\vect{w}}[j]\\
	\end{aligned}
\end{equation}

\autoref{eq:zkcnn:conv} becomes
\begin{equation}\label{eq:zkcnn:Ujk}
	\vect{U}[i][j] = \bar{\vect{u}}[n^2-1-sni- sj]
\end{equation}
To compute 1-D convolution using the fast Fourier transform (FFT) and inverse FFT (IFFT), $\bar{\vect{x}}, \bar{\vect{w}}$ are transformed to polynomials $\bar{\vect{x}}(\eta), \bar{\vect{w}}(\eta)$ with $\bar{\vect{x}}, \bar{\vect{w}}$ as coefficients, then $\bar{\vect{u}}(\eta)=\bar{\vect{x}}(\eta)\bar{\vect{w}}(\eta)$ by taking $\bar{\vect{u}}$ as the first $n^2$ coefficients. In zkCNN, the convolution  $\bar{\vect{x}} * \bar{\vect{w}}$ can be proven by 

\begin{equation*}
	\bar{\vect{u}} = \bar{\vect{x}}*\bar{\vect{w}} = \mathsf{IFFT}(\mathsf{FFT}(\bar{\vect{x}})\odot \mathsf{FFT}(\bar{\vect{w}}))
\end{equation*}
where $\odot$ represents the Hadamard product. Since the size of $\bar{\vect{x}}$ and $\bar{\vect{w}}$ are $n^2$, the proving time is $O(n^2)$, the verifier's time and proof size are $O(\log^2 n)$ given oracle access to the multilinear extensions of the input and the output.

We observe that in \autoref{eq:zkcnn:Ujk}, the majority of terms in $\bar{\vect{u}}$ are not the convolutional results when $s \geq 2$.
%
%
By applying our split technique to \autoref{eq:zkcnn:conv}, we show that the proving time for IFFT can be reduced by $s$ times. 
Specifically, we split $\vect{X}, \vect{W}$ for $s$ times respectively, such that

\begin{equation*}
	\begin{aligned}
		\vect{U}[i][j] = \sum_{k=1}^{s} \left[\sum_{u=0, v=0}^{w-1, w/s-1} \vect{X}[si+u][sj+sv+k]\cdot \vect{W}[u][2v+k] \right]
	\end{aligned}
\end{equation*}

Instead of creating $\bar{\vect{x}}$ and $\bar{\vect{w}}$ of size $n^2$, we transform $\vect{X}[si+u][sj+sv+k], \vect{W}[u, 2v+k]$ to $\bar{\vect{x}}^{(k)}, \bar{\vect{w}}^{(k)}$, respectively, following the same rule as in \autoref{eq:zkcnn:barXWU}. Then 

\begin{equation} \label{eq:zkcnn:barU}
	\begin{aligned}
		\bar{\vect{u}}^{(k)}[i] = \sum_{j=0}^i \bar{\vect{x}}^{(k)}[i-j]\cdot \bar{\vect{w}}^{(k)}[j]
	\end{aligned}
\end{equation}

The prover could use FFT to prove the correctness of \autoref{eq:zkcnn:barU} such that

\begin{equation*}
    \sum_{k=1}^{s} \bar{\vect{u}}^{(k)} = \mathsf{IFFT}\left(\sum_{k=1}^{s} (\mathsf{FFT}(\bar{\vect{x}}^{(k)}(\eta)) \odot \mathsf{FFT}(\bar{\vect{w}}^{(k)}(\eta)))\right)
\end{equation*}

By adopting the split technique to convolution layers in zkCNN, the proving time for the inverse FFT is reduced by $s$. 
To further demonstrate how it works,
we provide an example in \autoref{fig:zkcnnexample} when $n=4, w=2$, and $s=2$. 
As shown in \autoref{fig:zkcnnexample}, 
directly transforming the inputs and kernels results in the vectors of size $n^2=16$ (case {\circled[1]{\small \textsc{1}}}). 
Adopting the split technique reduces the dimension to $\frac{1}{2} n^2 = 8$ (case {\circled[1]{\small \textsc{2}}}).
Based on the zkCNN implementation \cite{code:zkcnn}, our experiments showed that adopting the split technique reduces the proving latency of IFFT in zkCNN from approximately 2 to 4 times in Lenet and VGG11 datasets (\autoref{fig:exp:zkcnn}).

\section{Security Proofs}\label{sec:proveMainTheorem}

\begin{proof}[Proof of \autoref{theorem:sys:security}]
	We argue the completeness, soundness, and zero-knowledge properties of our scheme as follows.
	
	\PP{Completeness}
	The circuit in $\sys.\Prover$ outputs 1 if $ y $ is the correct inference label of data sample $ \vect{x} $ by \autoref{alg:MLIP:DWTPCASVM} on MLIP parameters $ \mlparam $.
	The correctness of our protocol in \autoref{fig:sys:protocol} follows the correctness of the backend ZKP protocol by \autoref{theorem:spartan}.
	
	\PP{Soundness}
	{Let $C$ be the arithmetic circuit that represents the computation of MLIP with DWT, PCA, and SVM.} By the extractability of commitment used by the backend ZKP, there exists an extractor $ \Extor $ such that given $ \cm $, it extracts a witness $ w^* $ such that $ \cm = \zkp.\Commit(w^*,r, \pp) $  with overwhelming probability. 
	By the soundness of \zkmlip~in \autoref{def:sec:zkmlip}, if $ \cm = \zkmlip.\Commit(\mlparam,\pp,r) $ and $ \zkmlip.\Verifier(\cm,\vect{x},y,\pi,\pp) = 1 $ but $ y \ne \funcml(\mlparam, \vect{x}) $, then there are two scenarios:
	\begin{itemize}[leftmargin=*]
		\item Scenario 1: $ w^* = (\mlparam,\mathsf{aux}) $ satisfying to $C((\cm,\vect{x},y,\vect{r}');w^*) = 1$. 
		There are three cases for this to happen: 
		$(i)$ $ \mlparam $ is not the one committed to $ \cm $ but passing the verification for $ \cm $;
		$(ii)$ $ y $ is not the class label corresponding with the maximum predicted value among the auxiliary witnesses $(f^{(1)},\dots,f^{(s)}) \in \mathsf{aux}$ in \autoref{eq:svm:final}, but passing the max and permutation test;
		$(iii)$ Some witnesses in $ \mathsf{aux} $ are not valid, but passing the random linear combination test.
		The probability of the first case is negligible in $ \secparam $ due to the soundness of the commitment scheme used by the backend ZKP protocol.
		As $ \mathsf{Max} $ gadget relies on the permutation test, 
		its soundness error is negligible in $ \secparam $ 
		due to the soundness of the characteristic polynomial check, which achieves the probability of $ s/|\F| $ due to Schwartz-Zippel Lemma \cite{schwartz1980fast}.
		Finally, the soundness error of the random linear combination over a small number of constraints is negligible in $ \secparam $. 
		By the union bound, the probability that $ \Prover $ can generate such $ w^* $ is $\mathsf{negl}(\secparam)$.
		\item Scenario 2: $w^* = (\mlparam,\mathsf{aux})$ and $C((\cm,\vect{x},y,\vec{\alpha});w^*) = 0$. According to the soundness of the backend ZKP, given a commitment $ \cm^* $, the probability that $ \Adv $~can generate a proof $ \pi_w $ making $ \Verifier $ accept the incorrect witness is negligible in $ \secparam $.
	\end{itemize}
	In overall, the soundness of \sys~holds except with a negligible probability in $ \secparam $.

	\begin{figure}[!t]
		\small
		\begin{mdframed}
			\begin{Simulator}[Simulation of \autoref{pro:sys}] \label{sim:sys}
				Let $\lambda$ be the security parameter,  $\mathbb{F}$ be a finite field, $ \mlparam $ with $ n $ values. {Let $\pp \gets \sys.\Gen(1^\secparam)$}.
				\begin{itemize}
					\item $ \hat{\cm} \gets \Sim_1(n, r, \pp)$: $ \Sim_1 $ invokes $\Sim_\zkp$ to generate $\hat{\cm} = \Sim_\zkp(n, r, \pp)$ where $ r $ is randomness generated by $\Sim_\PC$.
					
					\item $(y,\pi) \gets \Sim_2^{\Adv}(\mlparam,\vect{x},\pp)$: $ \Sim_2 $ queries the oracle to get $y \gets \mathsf{DPS}(\mlparam,\vect{x})$. $\Sim_2$ shares all public input of $ C $ to $ \Sim_{\zkp} $ and invokes $ \cm_w \gets \Sim_{\zkp}.\Commit(\pp) $. 
					Upon receiving randomness $\vec{\alpha}$ from $ \Adv $, $\Sim_2$ invokes {$\pi \gets \Sim_{\zkp}.\Prover(C,(\hat{\cm},\vect{x},y,\vec{\alpha}),\pp)$}, and sends $\pi$ to $\Adv$. 
					
					\item $b \gets \Adv(\cm,\vect{x},y,\pi,\pp)$: Let $\cm = (\hat{\cm},\cm_w)$, wait $ \Adv $ for validation.
				\end{itemize}
			\end{Simulator}	
		\end{mdframed}\vspace{-.7em}
		\caption{Simulator of \autoref{pro:sys}.} \label{fig:sys:protocol:sim}
	\end{figure}
	
	\PP{Zero Knowledge}
	We construct a simulator for \autoref{pro:sys} in \autoref{fig:sys:protocol:sim} and show that the following hybrid game is indistinguishable.
	
	\begin{itemize}
		\item \textbf{Hybrid} $H_0$: $H_0$ behaves as the honest prover in \autoref{pro:sys}.
		\item \textbf{Hybrid} $H_1$: $H_1$ uses the real $ \sys.\Commit() $ in \autoref{pro:sys}, for the commitment phase, and invokes $\Sim$ to simulate the proving phase.
		\item \textbf{Hybrid} $H_2$: $ H_2 $ behaves as \autoref{sim:sys}.
	\end{itemize}
		
	Given the same commitment, the verifier cannot distinguish $H_0$ and $H_1$ due to the zero-knowledge property of the backend zero-knowledge protocol, given the same circuit $ C $ and public input. 
	If the verifier can distinguish $ H_1 $, and $ H_2 $, we can find a PPT adversary $ \Adv $ to distinguish whether a commitment of an MLIP with zero strings or not, which is contradictory with the hiding property of the underlying commitment scheme.
	Thus, the verifier cannot distinguish $ H_0 $ from $ H_2 $ by the hybrid, which completes the proof of zero-knowledge.
\end{proof}

\section{Proving Other SVM Kernels}\label{sec:other_kernel}
	Let $ c \in \F $ be the output of the kernel function. We present constraints for other SVM kernels as follows.
\begin{itemize}
	\item \emph{Laplace kernel. } $\phi_{\mathsf{la}}(\vect{x}_i, \vect{x}_j)  = e^{-\gamma' ||\vect{x}_i - \vect{x}_j||}$ can be proven with the following constraints

		\begin{equation}
		\begin{cases}
			b = -\gamma' ||\vect{x}_{i}-\vect{x}_{j}|| \\
			\mathsf{Exp}(c, e, b)
		\end{cases}
	\end{equation}

	where $ b \in \F $ is intermediate value.
	\item \emph{Sigmoid kernel. } $\phi_{\mathsf{sig}}(\vect{x}_i, \vect{x}_j)  = tanh[\alpha(\vect{x}_i^T \vect{x}_j)-\beta]$, where $\alpha, \beta >0$ are hyper-parameters, can be proven with following constraints
	%
	

			\begin{equation*}
		\begin{cases}
			b =  \alpha(\vect{x}_i^T \vect{x}_j)-\beta\\	
			\mathsf{Exp}(a_1, e, b)\\
			a_1\cdot a_2 = 1\\
			c\cdot(a_1 +a_2) = a_1 - a_2
		\end{cases}
	\end{equation*}
\end{itemize}
	where $b \in \F$ is the intermediate value, and $a_1, a_2 \in \F$ are auxiliary witnesses.

\section{Proving Deep Learning Techniques}\label{sec:proveDL}
	In this paper, we mainly focus on designing techniques to prove classical ML algorithms in zero-knowledge. However, we show that they can be used to prove some deep learning techniques as follows.
	
	\PP{Convolutional layers} A convolutional layer computes the dot product between an input vector $\vect{x} \in \F^n$ and a small kernel $\vect{k} \in \F^c$. 
	In the $i$th round, it computes the $i$th entry of the output such that
		$o[i] = \sum_{j=1}^c \vect{k}[j]\cdot \vect{x}[s(i-1)+j]$,
	where $s$ is the step between two rounds. 
	Our proposed technique can be applied to the convolutional layers w.r.t different settings of $s$.
	\begin{itemize}
		\item \emph{$s=1$. } This includes only addition and multiplication operations. Thus, the random linear combination can be applied to reduce the number of constraints, or other optimization techniques  \cite{liu2021zkcnn, lee2020vcnn} can be used.
		\item \emph{$s = 2$. } Our split technique in \autoref{sec:subsubsec:dwt} can be applied. Both the kernel and inputs are split into two parts, and a random linear combination can be performed.
		\item \emph{$s \geq 2$. } Our split technique can be extended when the step is greater than two. We first split $\vect{x}$ and $\vect{k}$ to $s$ parts, such that in the $l$th part, 
		
			\begin{equation*}
				o[i]^{(l)} = \Sigma_{i=1}^{c/s} \vect{k}[l+s(j-1)] \cdot \vect{x}[l+s(i+j-2)]
			\end{equation*}
			
		and $o[i]$ can be computed as 
		
		\begin{equation*}
			o[i] = \Sigma_{l=1}^{s} o[i]^{(l)}
		\end{equation*}
		
		Then the random linear combination can be utilized as described in \autoref{sec:subsubsec:dwt}.
	\end{itemize}
	
	\PP{Activation layers} Let $ c \in \F $ be the output of the activation function. 
	We show how to prove activation functions with our gadgets as follows.
	
	\begin{itemize}
		\item \emph{Sigmoid activation. } $f_{\mathsf{sig}}(x) = \frac{1}{1+e^{-x}}$ can be proven with following constraints

		\begin{equation*}
			\begin{cases}
				\mathsf{Exp}(a, e, x)\\
				a = (1 + a)\cdot c 
			\end{cases}
		\end{equation*}

	where $a \in \F$ is the auxiliary witness. 
	
		\item \emph{ReLU activation. } $f_{\mathsf{relu}}(x) = \mathsf{max}(x, 0)$ can be proven with $\mathsf{Max}(c,(x,0))$ gadget.

		\item \emph{Leaky ReLU activation. } $f_{\mathsf{lrelu}}(x) = \mathsf{max}(0.01x, x)$ can be proven as $\mathsf{Max}(c, (b, x))$ where $b = 0.01x $  is intermediate value.

		\item \emph{Tanh activation. } 
		$f_{\mathsf{tanh}}(x) = \frac{e^x - e^{-x}}{e^x + e^{-x}}$ can be proven with following constraints

				\begin{equation*}
			\begin{cases}
				\mathsf{Exp}(a_1, e, x)\\
				a_1\cdot a_2 = 1\\
				c\cdot(a_1+a_2) = a_1-a_2\\
			\end{cases}
		\end{equation*}

	\end{itemize}

	where  $a_1, a_2 \in \F$ are auxiliary witnesses.%
	
	\PP{Pooling layers} The max pooling layer $ c = \mathsf{max}(\vect{x})$ can be proven with $\mathsf{Max}(c,\vect{x})$ gadget.

\section{Mitigating Model Stealing Attacks} \label{sec:counterMEA}

As discussed, model stealing attacks \cite{tramer2016stealing, chandrasekaran2020exploring, jagielski2020high} aim to reconstruct the ML model from the inference result, 
given that the adversary has black box access to the model parameters.
%
To our knowledge, there is no general defense against these attacks beyond limiting the number of queries the client can make to the model \cite{jagielski2020high}.
We present several strategies that can mitigate these attacks, and, with some efforts, they can be integrated orthogonally into our scheme to protect the model privacy for both the inference result and the proof.

\PP{Limiting prediction information}The model holder can limit the output information by releasing class probabilities only for high-probabilities classes (e.g., top-5 in ImageNet dataset \cite{krizhevsky2012imagenet}) \cite{tramer2016stealing}, or only releasing the class labels \cite{tramer2016stealing, chandrasekaran2020exploring}. 
Limiting output information forces the adversary to query more, which permits the model holder to identify them by augmenting adversarial detection methods (see below) that analyze their behaviors against benign users. 
Tramer et al. \cite{tramer2016stealing} showed that by returning the class label without the confidence score (like \sys~currently offers), the number of required queries to extract the model increases by $50$-$100$ times. 
Thus, the model holder can increase the cost per query, thereby reducing the profit the adversary can make.

\PP{Adversarial detection}
Juuti et al. \cite{juuti2019prada} proposed an efficient method to detect whether the adversary is attempting to steal the model by analyzing the distribution of the adversary's queries against the normal (Gaussian) distribution.
Kesarwani et al. \cite{kesarwani2018model} proposed two performance metrics (e.g., the information gain and the coverage of the input space)
that quantify the rate of information the adversaries gained from the queries and are used to represent the status of the model extraction process.
Another approach is to embed watermark techniques so that if the adversary steals the model, the owner can detect and certify the stolen model \cite{272262, extraction:watermark:adi}.

\PP{Obfuscating prediction results}Several approaches suggest perturbing or adding noise to the prediction results to prevent the adversary from executing the (supervised) retraining process to reconstruct the model \cite{tramer2016stealing, chandrasekaran2020exploring, lee2018defending}.
This can be achieved with Differential Privacy to hide the decision boundary between prediction labels regardless of how many queries are executed by the adversary \cite{StealingAttack:Counter:DP:zheng2020protecting}.
Another approach is to poison the training objective of the adversary by actively perturbing the predictions without impacting the utility for benign users \cite{StealingAttack:Counter:DP:Orekondy2020Prediction}.

\section{Model Leakage in Proof of Inference without Zero Knowledge} \label{sec:ModelLeakageFromNonZKPoI}

We show how the proof of inference, without zero-knowledge, can leak model parameters.
Let $\vect{w}\in \F^{n}$ be the MLIP model parameters, $\vect{x} \in \F^{m}$ be the public inputs and outputs, and $s = \lceil \log n \rceil$. According to Spartan, our backend ZKP protocol, the secret parameter $\vect{z} = (\vect{x}, 1, \vect{w})$ is encoded as a function $Z(\cdot) : \{0,1\}^{s} \rightarrow \F$ that the low degree extension of it is a multilinear polynomial $\tilde{Z}(\vect{y})$, such that

	\begin{equation*}
		\tilde{Z}(\vect{y}) = \Sigma_{\vect{e}\in \{0,1\}^s} Z(\vect{e}) \cdot \prod_{i=1}^s(y_i \cdot e_i + (1-e_i)(1-y_i))
	\end{equation*}

To prove the satisfiability of the arithmetic circuits, both parties invoke two sumcheck protocols, where a dot-product-proof protocol \cite{wahby2018doubly} is applied to guarantee the zero-knowledge property. Suppose we do not have the zero-knowledge property, the sumcheck protocol would leak the information of the secret parameter $\vect{z}$. 
Specifically, in the first round of the sumcheck protocol, upon receiving a random challenge $\vect{r}_x\in \F^s$, $\Prover$ computes 
$v_A = \Sigma_{\vect{y} \in \{0,1\}^s}\tilde{A}(\vect{r}_x, \vect{y})\cdot \tilde{Z}(\vect{y})$,	
where $\tilde{A}: \F^{s} \times \F^{s} \rightarrow \F$ is a sparse multilinear polynomial, which is the low degree extension of matrix \vect{A} in R1CS. 
Therefore, once acquiring $v_A$, $\Verifier$ could compute the value of $\tilde{Z}(\vect{y})$, which contains private information of the model.
This demonstrates the importance of having zero-knowledge in the integrity proof to protect the model parameter privacy.

\end{document}